\documentclass{aa}
\usepackage{graphicx}
\usepackage{natbib}
\usepackage{txfonts}
\usepackage{epstopdf}
\usepackage[utf8]{inputenc}
\usepackage[english]{babel}
\bibpunct{(}{)}{;}{a}{}{,} 
\usepackage[dvipsnames]{xcolor}
\usepackage[colorlinks=true,citecolor=blue,linkcolor=magenta,urlcolor=Cerulean]{hyperref}
\def\apjs{ApJS}

\def\aap{Astron. Astrophys.}

\begin{document}
\title{Rotation periods from the inflection point in the power spectrum of stellar brightness variations: II.~The~Sun}
\titlerunning{GPS: II. The Sun}

\author{ E.M.~Amazo-G\'{o}mez \inst{1} \and A.I.~Shapiro \inst{1} \and S.K.~Solanki \inst{1,3} N.A.~Krivova \inst{1} \and G.~Kopp \inst{4} \and T.~Reinhold \inst{1} \and M.~Oshagh \inst{2} \and A.~Reiners \inst{2}}
\offprints{E.M.~Amazo-G\'{o}mez}

\institute{Max-Planck-Institut f\"{u}r Sonnensystemforschung, Justus-vonustus-von-Liebig-Weg 3, 37077, G\"{o}ttingen, Germany\\
\email{amazo@mps.mpg.de}
\and Georg-August Universit\"{a}t G\"{o}ttingen, Institut f\"{u}r Astrophysik, Friedrich-Hund-Platz 1, 37077 G\"{o}ttingen, Germany
\and School of Space Research, Kyung Hee University, Yongin, Gyeonggi 446-701, Korea
\and Laboratory for Atmospheric and Space Physics, 3665 Discovery Dr., Boulder, CO 80303, USA}
\authorrunning{E.~M.~Amazo-G\'omez et al.}

\date{Received ; accepted }

\abstract 
{Young and active stars generally have regular, almost sinusoidal, patterns of variability attributed to their rotation, while the majority of older and less active stars, including the Sun, have more complex and non-regular light-curves which do not have clear rotational-modulation signals. Consequently, the rotation periods have been successfully determined only for a small fraction of the Sun-like stars (mainly the active ones) observed by transit-based planet-hunting missions, such as CoRoT, Kepler, and TESS. This suggests that only a small fraction of such systems have been properly identified as solar-like analogs.}
{We apply a new method for determining rotation periods of low-activity stars, like the Sun. The method is based on calculating the gradient of the power spectrum (GPS) of stellar brightness variations and identifying a tell-tale inflection point in the spectrum. The rotation frequency is then proportional to the frequency of that inflection point. In this paper test this GPS method against available photometric records of the Sun.}
{We apply GPS, autocorrelation functions, Lomb-Scargle periodograms, and wavelet analyses to the total solar irradiance (TSI) time series obtained from the Total Irradiance Monitor (TIM) on the Solar Radiation and Climate Experiment (SORCE) and the Variability of solar IRradiance and Gravity Oscillations (VIRGO) experiment on the SOlar and Heliospheric Observatory (SoHO) missions. We analyse the performance of all methods at various levels of solar activity.}
{We show that the GPS method returns accurate values of solar rotation independently of the level of solar activity. In particular, it performs well during periods of high solar activity, when TSI variability displays an irregular pattern and other methods fail. Furthermore, we show that after analysing the light-curve skewness, the GPS method can give constraints on facular and spot contributions to brightness variability.} 
{Our results suggest that the GPS method can successfully determine the rotational periods of stars with both regular and non-regular light-curves.} 



\maketitle
\section{Introduction}\label{sec:intro}

\noindent A star's rotation period defines the action of the stellar dynamo, transport of magnetic flux through the convective zone, and its emergence on the stellar surface. (See \citealt{2010LRSP....7....3C} for a general review of dynamo theory, and \citealt{2014ApJ...794..144R}, \citealt{ 2018arXiv181006728I} for the relation between rotation period and activity.) Furthermore, for the unsaturated regime (i.e., when Rossby number is larger than~0.13, equivalent to rotation periods longer than 1? 10~days for solar-type stars) the magnetic activity of stellar chromospheres (as indicated by the Ca~II~H~\&~K emission core lines) and coronae (as indicated by the X-ray emission) monotonically increases with the decrease of the rotation period \citep{2003A&A...397..147P,2011ApJ...743...48W,2012LRSP....9....1R,2014ApJ...794..144R}. 

Recent studies indicate that the relationships between rotation period and coronal and chromospheric activity work not only for stars with a tachocline (the  transition region between the radiative core and convective envelope, see \citealt[]{ 1992A&A...265..106S}), but also for slowly-rotating fully convective stars \citep[see][]{2012AJ....143...93R,2016Natur.535..526W,2017ApJ...834...85N,2018MNRAS.479.2351W}. The rotation period thus appears to be a good proxy of the overall magnetic activity of a star.

Accurate measurements of rotation periods are important for understanding stellar evolution and for better calibration of the gyrochronology relationship \citep{1986ApJ...306L..37U,2003ApJ...586..464B}. The knowledge of the stellar rotation period helps distinguish the radial velocity jitter of a star from the planetary signal \cite[see, e.g.][]{2011A&A...525A.140D,2018ASSP...49..239O}. This is crucial for detection, as well for confirming Earth-size planets in ongoing and upcoming surveys, such as the ESPRESSO \citep[see][]{EXPRESSO} and PLATO missions \citep[see][]{2007CoAst.150..357R,2016AN....337..961R,2014ExA....38..249R}. 

Rotation periods can be determined from photometric observations thanks to the presence of magnetic features on stellar surfaces. Concentrations of strong localised magnetic fields emerge on the stellar surface and lead to the formation of photospheric magnetic features, such as bright faculae and dark spots \citep{2006RPPh...69..563S}. The transits of these magnetic features over the visible disk as the star rotates imprints particular patterns onto the observed light-curve. These patterns provide a means of tracing stellar rotation periods. We note that in the case of the Sun, such brightness variations are well understood \citep{2013ARA&A..51..311S,acp-13-3945-2013} and modern models can explain more than 96$\%$ of the variability of total solar irradiance (TSI, which is the spectrally-integrated solar radiative flux at one Astronomical Unit from the Sun) \citep[see][]{2014A&A...570A..85Y}.

Available methods for retrieving rotation periods from photometric time series, e.g.,  autocorrelation analysis or Lomb-Scargle periodograms, appeared to be very successful for determining periods of active stars with periodic patterns of variability. The transiting planet-hunting missions such as COROT, Kepler, and TESS, \citep{2003A&A...405.1137B,2010Sci...327..977B,2015JATIS...1a4003R} opened unprecedented possibilities for acquiring accurate high-precision photometric time-series of stars different from the Sun. The new data from these missions enabled studies of stellar magnetic activity on a completely new level. In particular, it has become possible to measure rotation periods for tens of thousands of stars \citep{Walkowicz2013,2013A&A...560A...4R,Garcia2014,2014ApJS..211...24M}. At the same time, the pattern of brightness variations in slow rotators such as the Sun is often quasi-periodic and even irregular. The irregularities are mainly caused by the short (in comparison to stellar-rotation period) lifetimes of magnetic features, such as starspots/sunspots, and a large degree of randomness in the time and position of their emergence on the stellar surface. This renders the determination of rotation periods for low activity stars very difficult. For example, \cite{2018arXiv180304971V} showed that rotation periods of about 80\% of stars in the Kepler field of view with near-solar effective temperature remain undetected. Consequently, the stars with known rotation periods represent only the tip of the iceberg of Sun-like stars. This can lead to biases in conclusions drawn based on the available surveys of stellar rotation periods. The relatively low efficiency of standard methods in detecting periods of stars with non-regular light curves might affect studies aimed at comparisons of solar and stellar variability. For example, solar variability appears to be normal when compared to main-sequence Kepler stars with near-solar effective temperatures \citep{basrietal2013, Reinhold_sub}. At the same time, when comparisons are limited to main-sequence stars with near-solar effective temperature and with {\it known} near-solar rotation periods, the solar variability appears to be anomalously low \citep{Reinhold_sub}. One possible explanation of such a paradox is the inability of standard methods to reliably detect rotation periods of stars with light curves similar to that of the Sun \citep[see also discussion in][]{Witzke2020}. Along the same line, \cite{2019A&A...621A..21R} proposed that biases in determining rotation periods might contribute to the explanation of a dearth of intermediate rotation periods observed in Kepler stars \citep[][]{McQuillan2013a, 2013A&A...560A...4R, 2014ApJS..211...24M, Davenport2017}.

In \cite{paperI} (hereafter, Paper~I), we proposed a new method for determining stellar rotation periods from the records of their photometric variability. The method is applicable to late-type stars but particularly beneficial for and aimed at stars with low activity and quasi-periodic irregularities in their light-curves. The rotation period is determined from the profile of the high-frequency tail (i.e., its part in between about a day and 5--10 days) of the smoothed wavelet power spectrum. For this work we used Paul wavelet of order six \citep[see][]{1998BAMS...79...61T}.  We identified the point where the gradient of the power spectrum~(GPS) plotted on a log-log scale (in other words, $d~(\ln P(\nu))/d(\ln\nu) $, where $P$ is power spectral density and $\nu$ is frequency) reaches its maximum value. This point corresponds to the inflection point, i.e., the concavity of the power spectrum plotted in the log-log scale changes sign there. In Paper~I we have shown that the period corresponding to the inflection point is connected to the rotation period of a star by a calibration factor which is a function of stellar effective temperature, metallicity, and activity. 
 
The main goal of the present study is to test and validate the method proposed in Paper~I against the Sun, which presents a perfect example of a star with an irregular pattern of variability but known rotation period. In particular, by validating the GPS method against the Sun, we show that it has the potential to reduce biases caused by the relative inefficiency of standard methods to determine rotation periods of low-activity stars. More specifically, we utilise the calibration factor between the position of the inflection point and rotation period corresponding to the solar case and apply the GPS method to the available photometric records of the Sun. Furthermore, we test the performance of the GPS method at various levels of solar activity.

In Sect.~\ref{sec:Methods}, we give a short overview of the available methods for determining rotation periods as well briefly describe the GPS method. (A more detailed discussion of this method is given in Paper~I.) In Sect.~\ref{sec:Solar_case}, we compare the performance of our method with that of other available methods in the exemplary case of the Sun. In Sect.~\ref{sect:Noise_and_sk}, we present the relationship between solar activity and skewness of its photometric light-curve. Our main results are summarised in Sect.~\ref{sect:summary}.
 
\section{Methods for determining stellar-rotation periods}\label{sec:Methods}
 
High-precision, high-cadence photometric time-series allow the determination of accurate stellar-rotation periods, and the scientific community has developed various methods for retrieving periodicities embedded in these data. Current methods include autocorrelation functions, Lomb-Scargle periodograms, wavelet power spectra, and very recent techniques based on the Gaussian processes.

The autocorrelation function~(ACF) method is based on the estimation of a degree of self-similarity in the light-curve over time. The time lags at which the degree of self-similarity peaks are assumed to correspond to the stellar-rotation period and its integer multiplets. This assumption is valid if magnetic features which cause photometric variability are stable over the stellar-rotation period. 

The ACF method has been used by \cite{2014ApJS..211...24M} to create the largest available catalog of rotation periods until now: the rotation periods were found for 34,030 (25.6\%) of the 133,030 main-sequence Kepler target stars (observed in the Kepler star field at the time of the publication). 
The ACF calculations presented in our study have been performed with the A$\_$CORRELATE IDL function.

The Lomb-Scargle periodogram gives the the power of a signal a certain frequency~\citep[see][]{1976Ap&SS..39..447L,1982ApJ...263..835S}. Here we use the generalised Lomb-Scargle~(GLS) version~v1.03, applying the formalism given by \cite{2009A&A...496..577Z}. The GLS method is widely used for retrieving periodicities from time-series, and is applicable to stellar light-curves with non-regular sampling. In studies aimed at the determination of the rotation period, the highest peak in the GLS is assumed to correspond to the rotation period~\citep[see][]{2013A&A...560A...4R}. 

Wavelet power spectra (hereafter, PS) transform analysis is beneficial for time-series that have a non-stationary signal at many different frequencies. It has been employed for determining stellar-rotation periods by~\cite{2009A&A...506...41G}. To perform PS, we use the WV$\_$CWT IDL function based on the 6th~order Paul wavelet, \citep[see][]{1992AnRFM..24..395F,1998BAMS...79...61T}.

Another technique, which is currently being actively developed, is inferring rotation periods with the help of Gaussian processes \cite[GP,][]{2012RSPTA.37110550R}. GP can be applied to detect a non-sinusoidal and quasi-periodic behaviour of the signal in light-curves. The GP regression has been extensively tested for analysing various time-series, and, in particular, for retrieving the periodic modulation from stellar radial-velocity  \citep{2015MNRAS.452.2269R} and photometric \citep{2018MNRAS.474.2094A} signals.  GP performance can be readily compared with other approaches for determining rotation periods in small stellar samples \citep[see, e.g., ][who applied various techniques to  HD~41284]{2019arXiv191111714F}.
GP calculations, however, demand significant computational resources, e.g.,  analysis of a typical Kepler light curve takes from several hours to 12~hours and longer \citep{2018MNRAS.474.2094A}. The time efficiency of the GP~algorithms is expected to significantly improve with the development of new methods \citep{2017AJ....154..220F,2017zndo...1048287F}. Thus, it might be interesting in future studies to compare rotation periods determined by the GP and GPS algorithms. Such a comparison is, however, out of the scope of the present study.

The methods described above have been very successful for determining rotation periods of active stars with periodic patterns of photometric variability. However, their performance deteriorates for slower-rotating and consequently less-active stars, and, in particular, for stars with a near-solar level of magnetic activity. In such stars, the pattern of variability is complex and non-regular due to the irregular emergence of magnetic features that manifest as spots and faculae. In particular, the dark spots can lead to prominent dips in brightness, and we expect that for stars with solar-activity levels, their decay time is comparable to or even less than the stellar-rotation period, as it is for the Sun \citep[see][]{Solanki_2006,2017NatAs...1..612S}. Furthermore, \cite{2017NatAs...1..612S} showed that in the solar case, the superposition of bright facular and dark spot signatures might lead to a disappearance of the rotation peak from the power spectrum of solar brightness variations. This agrees with many studies \citep[see][]{2015MNRAS.450.3211A,2015ApJS..221...18H,2018IAUS..335....7H} which have found that the rotation period of the Sun can be reliably determined only during the periods of low solar activity. We note that during these periods, the solar variability is caused by long-lived facular features, and the spot contribution is small.  

Recently, \cite{2015ApJS..221...18H,2018IAUS..335....7H} analysed solar and stellar brightness using GLS and introduced two indicators: one describing the degree of periodicity in the light-curve, and the other describing the amplitude of the photometric variability. They found that light-curve periodicities of the Kepler stars are generally stronger during high-activity times. In contrast, solar light-curves are more periodic during phases of low activity. By applying GLS to the TSI time-series, they could determine the solar-rotation period only during periods of low solar activity.  

In Paper~I, we employed the SATIRE approach \citep[where SATIRE stands for Spectral and Total Irradiance Reconstruction, see][]{2000A&A...353..380F,2011A&A...529A..81K} to simulate light-curves of stars with solar effective temperature and metallicity for various cases of magnetic-feature emergence and evolution. We found that the profile of the wavelet power spectrum around the rotation period strongly depends on the decay time of magnetic features and the ratio between coverage of the disk by faculae and spots. At the same time, the high-frequency tail of the power spectrum, particularly the portion with periods between about a day and a quarter of the rotation period, remains much more stable, and mainly depends on the rotation period. To quantitatively characterise the high-frequency tail of the power spectrum, we calculated the ratios $R_k$ between the power spectral density, $P(\nu)$, at two adjacent frequency grid points: $R_k\equiv~P(\nu_{k+1})/P(\nu_k)$. For a frequency grid equidistant in the logarithmic scale (i.e., with a constant value of $\Delta \nu/\nu$), this ratio represents the gradient of the power spectrum plotted on a log-log scale. Consequently, the maxima of the $R_k$ values correspond to the positions of the inflection points, where the concavity of the power spectrum plotted on a log-log scale changes sign. Hereafter, following Paper~I, we will refer to the $R_k$ values as the gradient of the power spectrum, GPS.

In Paper~I, we found that the position of the high-frequency inflection point (i.e., the inflection point~(IP) with highest frequency, see discussion below) is related to the stellar-rotation period by a certain calibration factor defined as: $\alpha=HFIP/P_{\rm rot}$, where HFIP is a period corresponding to the high-frequency inflection point and $P_{\rm rot}$ is the rotation period. This allowed us to propose a new method for determining the rotation period: calculate the position of the inflection point and scale it with an appropriate value for the factor $\alpha$ to determine the rotation period.

The choice of the scaling factor $\alpha$ is one of the most delicate steps in the GPS method and also one of the main sources of its uncertainty. In Paper~I, we found that the value of $\alpha$ is fairly insensitive to the parameters describing the evolution of magnetic features, and, in particular, to the decay time of spots. It shows a stronger dependence on the area ratio between the facular and spot components of active regions at the time of maximum area, $S_{fac}/S_{spot}$. We have estimated that over the 2010?2014 period, the solar $S_{fac}/S_{spot}$ value was about~3 (see Appendix~B of Paper~I). For the calculations presented in  this paper we adopt $\alpha_{\rm Sun}\pm~2\sigma=0.158\pm0.014$, corresponding to $S_{fac}/S_{spot}=3$ and the spot decay rate of 25~MSH/day. Such a value of spot decay time is intermediate between the 10~MSH/day estimate by \cite{Waldmeier1955} and 41~MSH/day estimate by \cite{decay2}. 

The uncertainty of the $\alpha$~value is computed in Paper~I as the standard deviation between positions of inflection points corresponding to different realisations of spot emergences (but the same values of the $S_{fac}/S_{spot}$ ratio and spot decay rates). We stress that when applied to other stars, the uncertainty of our method will be significantly larger since neither $S_{fac}/S_{spot}$ value nor spot decay rates are known a priori. For example, in Paper~I we estimated that the internal uncertainty of our method is 25\%. Its main advantage, however, is that, in contrast to other methods, it is applicable to inactive stars with irregular patterns of brightness variability. 

In summary, we calculate the rotation period and its uncertainty as: 
\begin{equation}\label{eq1}
    P_{\rm rot}~\pm~2 \sigma_P=~\frac{HFIP}{(\alpha_{\rm Sun}~\pm~2\sigma_{\alpha})},
\end{equation}

\noindent where HFIP represents the value of the high-frequency inflection point in the power spectrum of brightness variations. 
 
\section{Validation of GPS method for the solar case}\label{sec:Solar_case}

In the present work, we evaluate the GPS method for determining solar rotation period against available records of its brightness variation. We consider the performance of the method at different levels of solar activity using two TSI data sets. 

\subsection{Records of total solar irradiance, TSI}

\noindent Driven by the interest from the climate community, the solar irradiance has been measured by various space radiometric instruments and reconstructed with many models, see, e.g., reviews by \citep{acp-13-3945-2013, 2013ARA&A..51..311S}.

Among all available solar-irradiance records, the TSI time-series are the most accurate and possess the longest time coverage \citep[see e.g.,  review by][]{2014JSWSC...4A..14K}. In this context, we have opted for testing the performance of the GPS method against the TSI time-series and focused on what are generally considered to be the two most reliable of the available data sets: one obtained by the Variability of solar IRradiance and Gravity Oscillations (VIRGO) experiment on the ESA/NASA SOlar and Heliospheric Observatory (SoHO) Mission \citep[see][]{1997SoPh..175..267F}, and another obtained by the Total Irradiance Monitor (TIM) on the Solar Radiation and Climate Experiment (SORCE) \citep[see][]{2005SoPh..230...91K,2005SoPh..230..111K,2005SoPh..230..129K}.

VIRGO provides more than 21~years of continuous high-precision, high-stability, and high-accuracy TSI measurements. Our analysis is based on the last update available at the beginning of this work, version~6.4$: 6\_005\_1705$, level~2.0 VIRGO/PMO6V observations from January~1996 until June~2017 with a cadence of 1~data point per hour~\footnote{ \href{https://www.pmodwrc.ch/forschung-entwicklung/sonnenphysik/tsi}{https://www.pmodwrc.ch/forschung-entwicklung/sonnenphysik/tsi-composite/} }. The data are available via ftp~\footnote{ \href{ftp://ftp.pmodwrc.ch/pub/data/irradiance/virgo/virgo.html}{ftp://ftp.pmodwrc.ch/pub/data/irradiance/virgo/virgo.html}}. 

The TIM data used are version 17, level 3.0, available from 
February 25, 2003,\footnote{ \href{http://lasp.colorado.edu/home/sorce/data/tsi-data/tim-tsi-release-notes/}{http://lasp.colorado.edu/home/sorce/data/tsi-data/tim-tsi-release-notes/}}~\footnote{ \href{http://lasp.colorado.edu/home/sorce/data/}{http://lasp.colorado.edu/home/sorce/data/}}. While TIM data cover a shorter time interval than VIRGO, they have a lower noise level \citep{2014JSWSC...4A..14K}, which is particularly important for our analysis of TSI variations during the minimum of solar activity. Here we use an average version of TSI with a cadence of 1~data point per~1.6 hours obtained by averaging over a full orbit of the spacecraft.~\footnote{Personal communication}.
 
\subsection{Brightness signature of spot transits}\label{subsec:Spot_signature}

\begin{figure*}[!ht]
\centering
\includegraphics[trim={0 0 0 0cm},clip,width=0.8\textwidth]{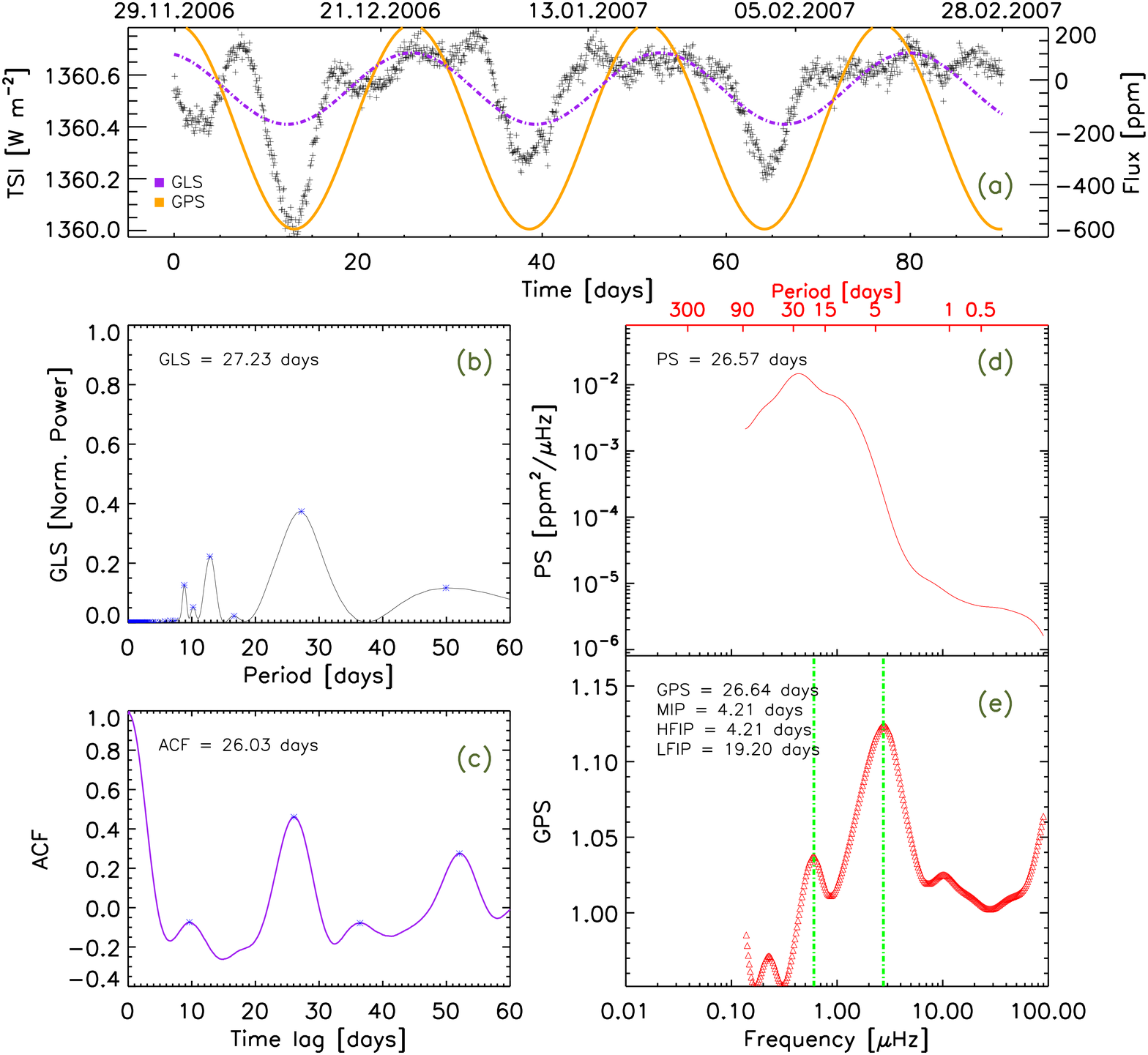}
\caption{Example of the spot-dominated total solar irradiance (TSI) variability. panel~(a) shows SORCE/TIM measurements from 28-Nov-2006 to 28-Feb-2007. Purple and orange curves indicate sine-wave functions with periods of 27.3~d (solar-rotation period deduced with the generalised Lomb Scargle periodogram method, GLS) and 26.6~d (solar-rotation period deduced with the gradient of power spectra method, GPS). Panels~(b) and~(c) show the corresponding GLS periodogram and autocorrelation function, ACF, respectively. Panel~(d) shows the global wavelet power spectrum, PS, calculated with the 6th~order Paul wavelet. Panel~(e) shows the gradient of the power spectrum plotted in panel~(d). Blue asterisk signs in panels~(b) and~(c) represent positions of the peaks in the GLS and ACF. Green dotted lines in panel~(e) indicate the high- and low-frequency inflection points.}
\label{Fig1}
\end{figure*}

In this section we analyse the performance of the GPS method during a 90-day time interval when TSI variability was generated by three consecutive spot transits. Sunspot transits imprint characteristic signals on the TSI, diminishing the observed solar brightness. Transits of well-resolved sunspots have been recorded by TIM/SORCE from December~2006 to February~2007 (see panel~(a) of Figure~\ref{Fig1}). One can see a clear signal from spots transiting the solar disk with a recognisable "V-like" shape brought about by the combination of foreshortening effects, the Wilson depression (i.e., spot umbrae are slightly depressed from the photospheric optical-depth unity level due to higher transparency of the spot atmosphere compared to the quiet photosphere (see \citealt{1965ApJ...142..773W})), and the center-to-limb dependence of spot-intensity contrast. 

Interestingly, the time separation between consecutive transits is very close to the solar-rotation period (compare sine functions and TSI time-series in Figure~\ref{Fig1}a). A fluctuating behaviour of transit amplitude (i.e., decrease of the amplitude from first to second transit and increase of the amplitude from second to third transit) suggests that we are observing transits of different spots, even though the emergence of all three spots occurred at the same location on the solar surface. We confirm this by comparing simultaneous records of the MDI intensity maps and TSI\footnote{see 2006\_TSI\_Movie.mp4 at:\\ $\href{ftp://laspftp.colorado.edu/3month/kopp/}{ftp://laspftp.colorado.edu/3month/kopp/}$}. Such a nesting of spot emergence \citep{1990SoPh..129..221B,1994ApJ...422..883G} affects the power spectrum of brightness variations exactly as the increase of the lifetime of the magnetic features. In particular, it makes light-curves more regular and helps to determine rotation period.

We perform a comparison between the GLS, ACF, PS, and GPS methods for the considered epoch of spot transits. In Figure~\ref{Fig1}b, we plot the results of the GLS analysis. A prominent peak is clearly observed at 27.2~days, which is very close to the {\it synodic} Carrington rotation period of 27.27~days. This peak has a power of 0.37, where GLS is normalised to unity. A second peak appears at 13.1~days with a normalised power of 0.23, which is a harmonic of the solar-rotation period. A large value of the normalised power in the rotation peak is not surprising: the corresponding light-curve has a clear periodicity, which results in a clear peak in the power spectrum. The purple dashed line in Figure~\ref{Fig1}a fits a sine wave with period determined from the GLS method.

Figure~\ref{Fig1}c shows the ACF analysis. Two clear peaks are visible at 26.0 and 52.0~days, having amplitudes of about 0.5 and 0.3, respectively. PS analysis is shown in Figure~\ref{Fig1}d. We use a Paul wavelet basis function with order m=6 to calculate the global wavelet power spectrum. PS show a pronounced peak around 26.6~days. One can also see a shoulder-like feature at about 13~days that is generated due to the ingress and egress of magnetic features over the visible solar disk. The contrast differences between faculae and spots transiting the limbs and the center will imprint a characteristic pattern in the light-curve, detected by the implementation of the GPS. 

Figure~\ref{Fig1}e shows the GPS profile with two well-defined peaks that correspond to the inflection points in the power spectrum. The low-frequency inflection point is at 19.2~days and is associated with the transition from the rotation peak to high-frequency modulations forming a shoulder-like feature. The high-frequency inflection point is at 4.21~days and corresponds to the transition from the shoulder-like feature to the plateau at higher frequencies. For the considered period of brightness variations due to spots, the high-frequency point also corresponds to the maximum value of the GPS. In other words, the gradient is larger at the high-frequency point than at the low-frequency point.

In Sect.~\ref{subsec:TSIasKepler}, we show that, in agreement with Paper~I, the high-frequency inflection point is present in the PS even when the rotation peak is absent. Furthermore, the location of the high-frequency inflection point is stable independently of the presence or absence of the rotation peak.

Based on the location of the high-frequency inflection point at 4.21~days and using Eq.~(\ref{eq1}) with the calibration factor $\alpha_{\rm Sun}=0.158$ and its 2$\sigma$ uncertainty of 0.014 we compute the solar-rotation period: $P_{\rm rot}=26.6^{+2.2}_{-2.6}$~days. We use the 26.6-day value for plotting the orange sine curve in Figure.~\ref{Fig1}a. This rotation period is in reasonably good agreement with the Carrington period given the relatively short length of the time-series.

In summary, all methods applied above allow reasonably accurate determinations of the rotation period for the considered period of pseudo-isolated transits of sunspots.

\subsection{Brightness signature of facular feature transits}\label{subsec:fac_signature}

\begin{figure*}[!ht]
\centering
\includegraphics[trim={0 0 0 0cm},clip,width=0.8\textwidth]{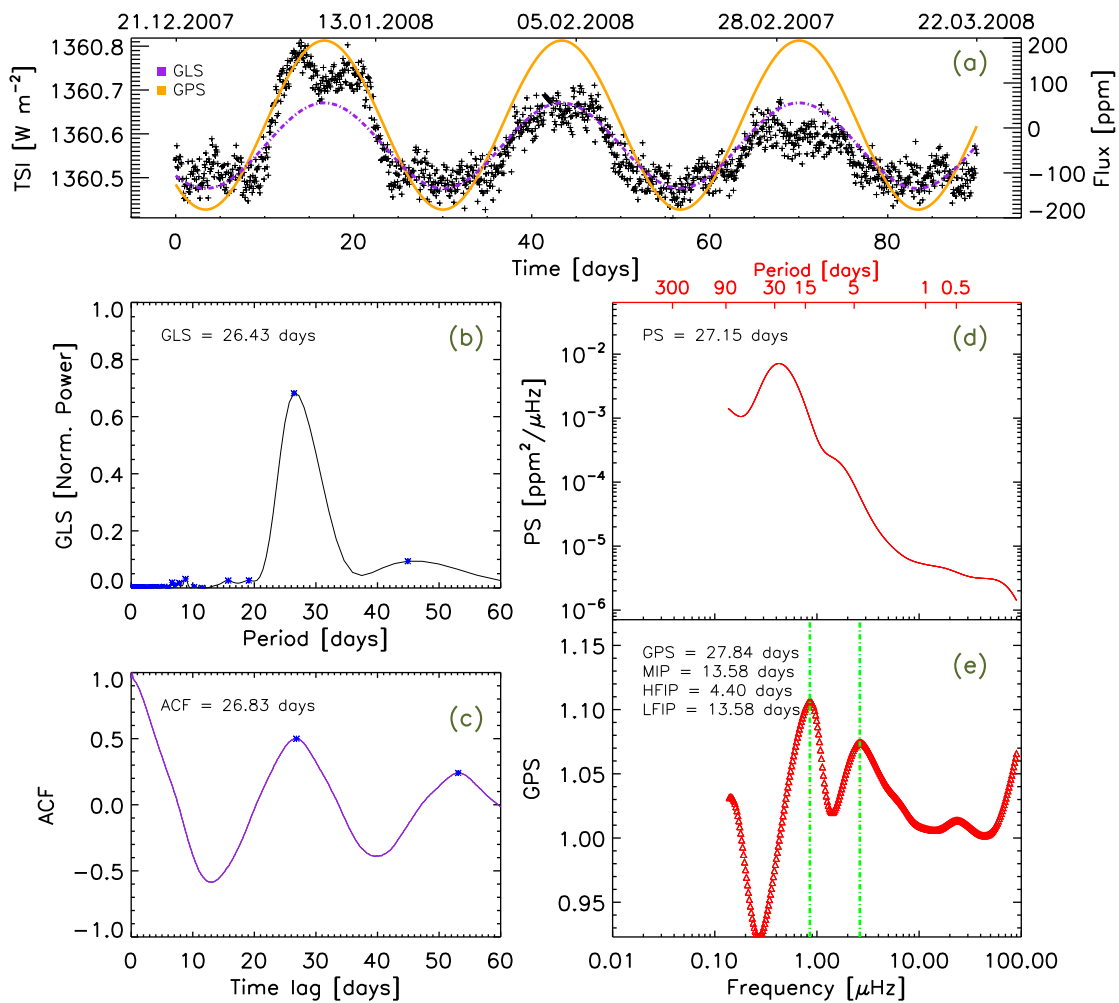}
\caption{The same as Figure~\ref{Fig1} but for the time interval of faculae-dominated TSI variability from 21-Dec-2007 to 22-Mar-2008. The purple dashed curve in panel~(a) represents a sine wave function with a period of 26.4~d corresponding to the solar rotation period deduced with the GLS method. The orange curve in panel~(a) is a sinusoidal function with a period of 27.8~d corresponding to the solar rotation period deduced with the GPS method.}
\label{Fig2}
\end{figure*}

\noindent Here we perform a rotation period analysis of an interval of time between December~2007 and March~2008 (see Figure~\ref{Fig2}) when the TSI variability was dominated by three consecutive transits of a facular feature.
 
The simultaneous analysis of MDI intensitigram, magnetograms, and TSI records \footnote{See 2007\_TSI\_Movie.mp4 at:\\ $\href{https://spot.colorado.edu/~koppg/TSI/}{https://spot.colorado.edu/~koppg/TSI/}$}\footnote{$\href{ftp://laspftp.colorado.edu/3month/kopp/}{ftp://laspftp.colorado.edu/3month/kopp/}$}shows that variability is brought about by a single large facular feature, which size is decreasing from transit to transit. We note the difference with the spot case described in Sect.~\ref{subsec:Spot_signature} where the three consecutive transits were caused by the nesting effect. Such a behaviour is consistent with the fact that lifetimes of facular features are significantly larger than that of the spots. 

The transit of a facular feature has a characteristic double-peak "M-like" profile, see Figure~\ref{Fig2}a. This "M-like" profile can be explained by the increase of facular contrast towards the limb. Such an increase partly compensates the foreshortening effect and, consequently, the maximum brightness occur when the facular feature is observed at an intermediate disk position.  

Figure~\ref{Fig2}b shows a GLS analysis of the considered time interval. A prominent peak is seen at 26.4~days with a normalised power of 0.68. The purple dash line in Figure~\ref{Fig2}a fits to the light-curve with a sine wave with a corresponding period. Figure~\ref{Fig2}c shows the corresponding ACF analysis. A clear series of peaks with time lag 26.8~days are observed. 

The PS analysis shows a peak at 27.2~days (see Figure~\ref{Fig2}d). Similarly to the case of spot transits one can see a shoulder-like feature, but it is shifted towards higher frequencies in comparison to the spot case. Such a shift of the shoulder-like feature is explained by the M-like profile of a facular transit in the light-curve. It leads to the enhanced variability on timescales shorter than the half of the rotation period and consequently shifts the shoulder-like feature.

Figure~\ref{Fig2}e shows the GPS profile with two clearly visible peaks that corresponds to the inflection points of the PS. As in the case of the variability brought about by sunspots there are two inflection points, one is closer to the rotation period peak and another to the shoulder-like feature. We note that for the faculae-dominated case, in contrast to the spot-dominated case, the maximum inflection point (i.e., the inflection point with highest value of the GPS) corresponds to the low frequency inflection point.  

The high-frequency inflection point is located at 4.40~days, which is very close to the location of the inflection point during the spot-dominated regime of the variability (4.21~days, see Sect.~\ref{subsec:Spot_signature}). Applying the calibration factor $\alpha_{\rm Sun}$ one can see that the value given by the GPS method for the solar rotation period is $P_{\rm rot}=27.8^{+2.3}_{-2.7}$~days. The orange line in Figure~\ref{Fig2}a fits a sinusoidal function with a period of 27.8~days as returned by the GPS analysis. The amplitude of the orange curve is given by the maximum amplitude of the wavelet. 

The analysis performed in this Section and in Sect.~\ref{subsec:Spot_signature} shows that when solar brightness variability is attributed to the periodic transits of spots and faculae all four methods (GLS, ACF, PS, and GPS) can accurately retrieve solar rotation period. However, the recurrent transits of one spot group or nested spots (as plotted in Figure~\ref{Fig1}a) are rare. Since faculae last significantly longer than spots, the recurrent transits of the same facular features are much more common than that of spots. However, faculae dominate the solar TSI light-curve on rotational timescales only at activity minimum (when no large spots are present, although on the solar cycle timescale, faculae play a dominant role). 

Most of the time solar brightness variations are brought about by the combination of magnetic features coming with random phases. As we will show below this strongly affects the performance of the GLS, ACF, and PS methods but has a much weaker effect on the GPS method.

\subsection{Analysis of the entire data-set}\label{Sub:Analysis of the entire data-set}

\begin{figure*}[!ht]
\centering
\includegraphics[trim={0 0 0 0cm},clip,width=0.8\textwidth]{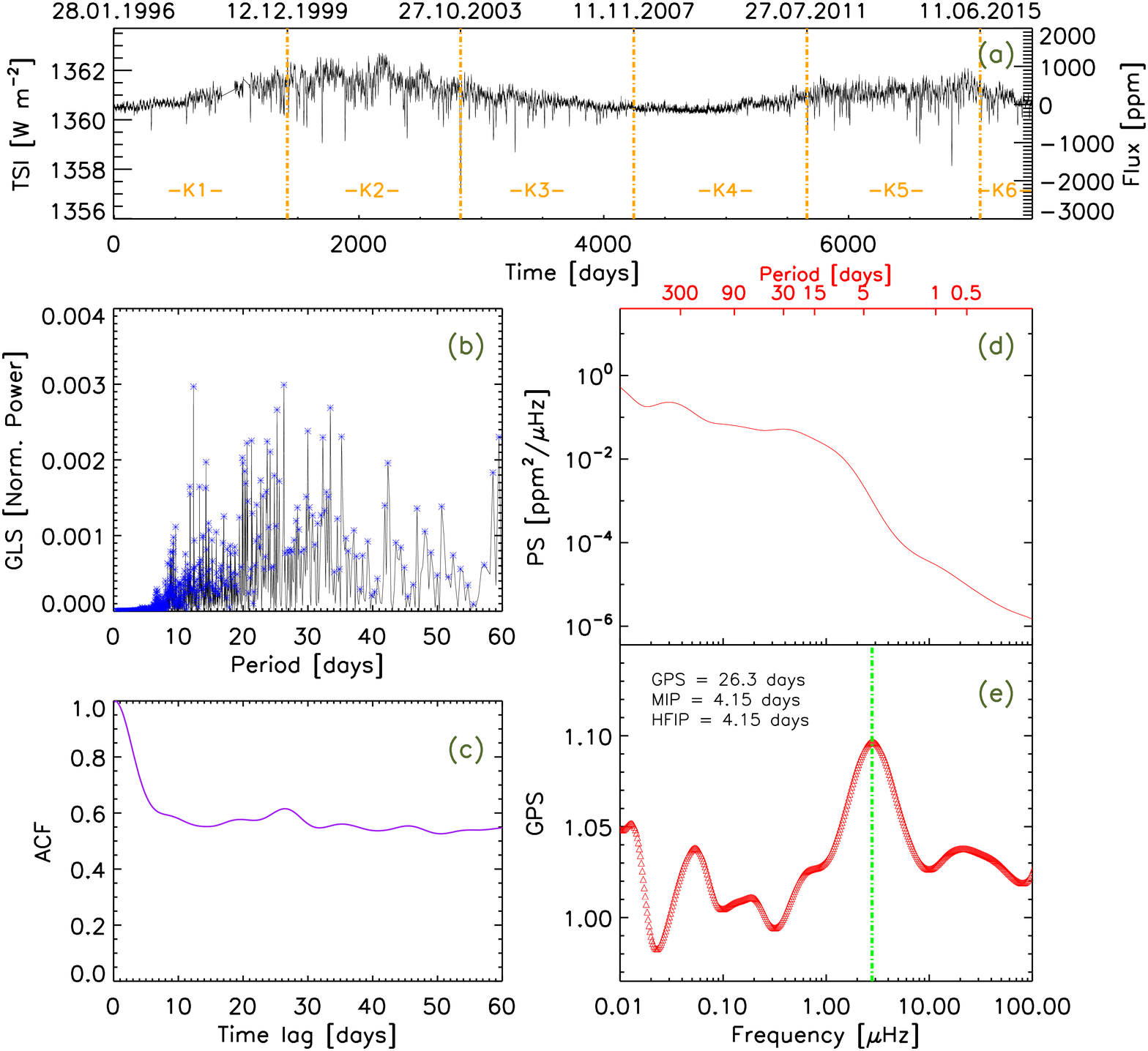}
\caption{The same as Figure~\ref{Fig1} but for 21~years [1996.01.28--2017.05.23] of TSI data from SoHO/VIRGO. In total 7787~days are considered. This corresponds to about 186889~data-points at an hourly cadence. The GLS, ACF, and PS methods do not show a clear signal of the rotation period. GPS shows a prominent peak at 4.15~days, resulting in a solar rotation period value of about 26.3~days (see text for details). Orange vertical lines and marks in panel~(a) represent a splitting of the VIRGO data-set into Kepler-like time-span $K_{n}$, each one representing the full lifetimes of the Kepler mission, around 4~years. (see Sect.~\ref{subsec:TSIasKepler}).} 
\label{Fig3}
\end{figure*}

\begin{figure*}[!ht]
\centering
\includegraphics[trim={0 0 0 0cm},clip,width=0.8\textwidth]{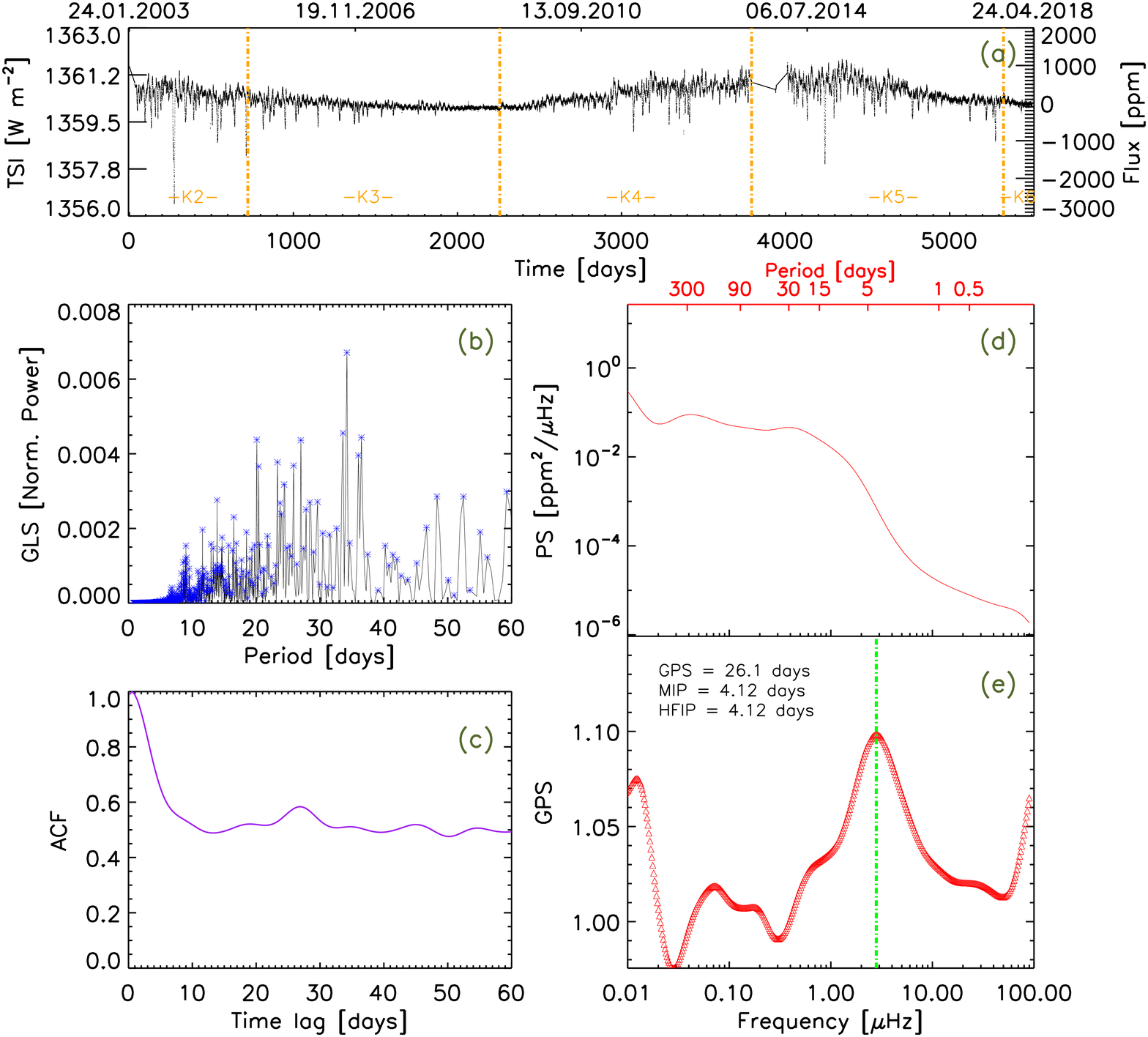}
\caption{The same as Figure~\ref{Fig1} but for 15~years [2003.01.24--2018.04.24] of TSI data from SORCE/TIM. In total 5508~days are considered. This corresponds to about 82632~data-points at a cadence of 1.6~hours. Orange lines represent the same Kepler-like time-span $K_{n}$ than shown in Figure~\ref{Fig3}. The GLS, ACF, and PS methods do not show a definitive detection signal of the rotation period. GPS method shows a prominent peak at 4.12~days corresponding to the solar rotation period value of 26.1~days.}
\label{Fig4}
\end{figure*}

In Sect.~\ref{subsec:Spot_signature}~and~\ref{subsec:fac_signature} we considered relatively short intervals of time when solar variability is dominated by either spot or facular components. In this section we study whether the solar rotation period can be reliably retrieved in a more general case when contributions from faculae and spots are entangled. For this, we consider 21~years of TSI data from SoHO/VIRGO (see Figure~\ref{Fig3}) and 15~years of TSI SORCE/TIM data (see Figure~\ref{Fig4}) and apply GLS, ACF, PS, and GPS methods to the entire time-series.

We plot GLS periodograms for VIRGO and TIM data in panels~(b) of Figs.~\ref{Fig3}~and~\ref{Fig4}, respectively. One can see that none of the two periodograms contain a sufficiently strong peak to provide a clear indication of periodicity. Instead they contain a series of peaks with similar and relatively small power: up to 0.003 for the VIRGO TSI and 0.007 for the TIM TSI. This is about two orders of magnitude lower than the normalised power obtained for pseudo-isolated spots and facular cases considered in Sects.~\ref{subsec:Spot_signature}~and~\ref{subsec:fac_signature}. 
Consequently, the GLS method does not allow a definitive detection of the rotation period when the entire TIM and VIRGO time-series are considered. The ACF of the VIRGO and TIM TSI time-series are shown in panels~(c) of Figs.~\ref{Fig3}~\&~\ref{Fig4}. 

Although we can appreciate small peaks at the expected location of the rotation period in both ACFs cases, the significance of the maxima with the corresponding time lag needed for the identification of the rotation period yields only a marginal detection. Panels~(d) in Figures~\ref{Fig3}~and~\ref{Fig4} display the PS analysis of the data. One can see that instead of the peak at the rotation period, one can observe a plateau region for both data-sets. 

All in all the, GLS, and PS methods cannot detect the solar rotation period, and ACF give us a marginal detection when long sets of TSI data are considered. This is in line with the result of \cite{2017NatAs...1..612S} who showed that the superposition of facular and spot contributions to solar variability can significantly decrease the rotation signal.

The GPS profiles of VIRGO and TIM time-series are given in Panels~(e) of Figures~\ref{Fig3}~ and~\ref{Fig4}. One can see that both profiles display conspicuous high frequency inflection points. They are located at 4.15 and 4.12~days for VIRGO and TIM data, respectively. Consequently, we retrieve a rotation period for the Sun: $P_{\rm rot}=26.3^{+2.1}_{-2.5}$~days, and $P_{\rm rot}=26.1^{+2.1}_{-2.5}$~days, for the VIRGO and TIM data, respectively. These P$_{\rm rot}$ values agree within the error bars with the solar {\it synodic} Carrington rotation period of 27.27~days, as well with the solar equatorial {\it synodic} rotation period for a fixed feature of 26.24~days. Consequently, the GPS method allows a proper determination of solar rotation period over timescales in comparison with other traditional approaches.

\subsection{The solar variability in 90-day quarters}\label{subsec:TSIasKepler}

Here we consider the exemplary case of the Sun as it would be observed over the same time-span as the Kepler stellar data-set. The Kepler telescope reoriented itself every 90~days, thus introducing discontinuities into the light-curves. To mimic the observational routine of the Kepler telescope, we segment the entire VIRGO and TIM data-sets into 86 and 60~quarters of 90-day duration, with a cadence of 1.0 and 1.6~hours, respectively. The cadence of solar observations are close to regular long cadence Kepler observations of 29.4~min. In this section we analyse the stability of the rotation signal from quarter to quarter, using the four methods described above. 

\begin{figure*}[!ht]
\centering 
\includegraphics[trim={0 0 0 0cm},clip,width=1.0\textwidth]{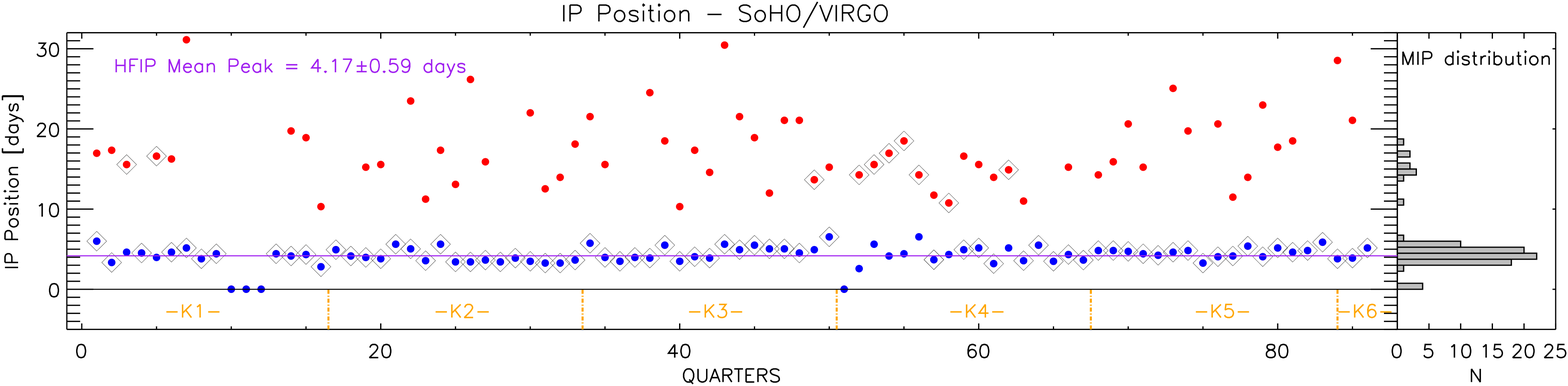}\hspace{0.0cm}
\includegraphics[trim={0 0 0 0cm},clip,width=1.0\textwidth]{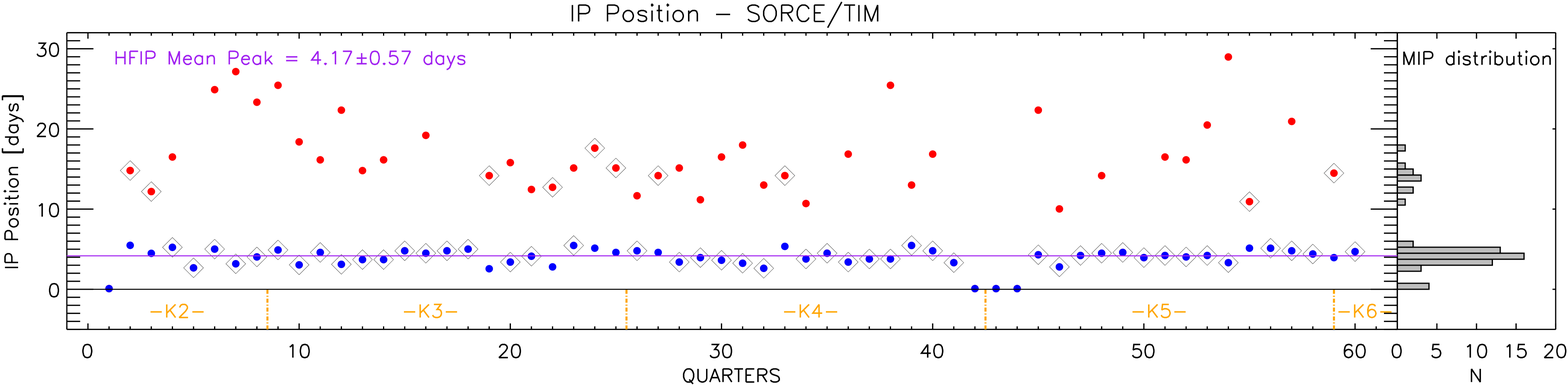}\hspace{0.0cm}
\caption{Top-Left panel: Positions of the inflection points for the 86 (90-day) quarters of the VIRGO data. Red dots represent the low frequency inflection points (LFIP), blue dots represent the high frequency inflection points (HFIP), black diamonds indicate inflection points with maximum GPS value (see text). Top-Right panel: Distribution of the maximum inflection point positions. Bottom panels: the same as top panels but for 60~quarters of 90-days using the TIM data. Orange lines in left panels indicate splitting of the VIRGO and TIM data-sets into Kepler-like time-span~$K_n$ as in Figure~\ref{Fig3}.}
\label{Fig5}
\end{figure*}

The inflection points described by the GPS method in all quarters are shown in 
Figure~\ref{Fig5} for the VIRGO (top panel) and TIM (bottom panel) data. We note that quarters 10, 12, 13 and 52 in the VIRGO data-set are affected by the lack of data due to spacecraft and instrument failures. TIM data quarters number 1, 43, 44, and 45 contain long gaps, some of them larger than 5~days. For the three recent quarters this is because of the failing SORCE battery. 

We note that power spectra of some quarters have more than one inflection point with considerable amplitude (like in the time intervals shown in Figs.~\ref{Fig1}~and~\ref{Fig2}). We analyse the two highest inflection points in the GPS and describe these in terms of its frequency location. The inflection point towards the highest frequency, HFIP, is represented by blue dots, in Figure~\ref{Fig5}. The inflection point towards lower frequencies, LFIP, is represented by red dots. In Figure~\ref{Fig5} for each quarter open black diamonds surround the inflection point corresponding to the maximum amplitude of the gradient of the power spectra, MIP. In most of the cases these maximum peaks are simultaneously the high frequency inflection points (as in the case of the variability brought about by spots, see Figure~\ref{Fig1}), marked with open black diamonds over-plotted over the blue dots. Similarly, there are several quarters when the maximum value of the gradient corresponds to the low frequency inflection point (like in the case of the variability brought about by faculae, as we explain before, see Figure~\ref{Fig2}). These quarters are associated with low solar activity and with TSI variability being dominated by faculae. Interestingly, the high frequency inflection points are still present in such quarters (even though they no longer correspond to the maximum amplitude of the gradient). In Figure~\ref{Fig5} we indicate when the maximum GPS amplitude corresponds to the low frequency inflection points with open black diamonds over-plotted over red dots.

Figure~\ref{Fig5}  demonstrates that positions of the inflection points are stable and the mean value of positions over all considered quarters are basically the same for the VIRGO and TIM data. We construe this as the prove that the GPS method works for the Sun. 

Having shown that the position of the inflection point is stable for the Sun, we could then use the Sun to calibrate the GPS method and apply $\alpha_{\rm Sun}$ value to other Sun-like stars, independently of the simulations presented in Paper~I. At the same time, it is reassuring to see that the theoretical $\alpha_{\rm Sun}$ value found in Paper~I leads to a reasonable value of the solar rotation period. Indeed, after applying the $\alpha_{\rm Sun}$ calibration factor from Paper~I to the position of the inflection point at 4.17~$\pm$~0.59~d for VIRGO data and 4.17~$\pm$~0.59~d for TIM data (where 0.59~d and 0.57~d corresponds to 2$\sigma$ of the observed distribution of values for all quarters) we obtain 26.4$~\pm~$3.7~d and 26.4$~\pm~$3.6~d, respectively.

We note that the uncertainty of the rotation period is calculated here in a different way than in Sects.~\ref{subsec:Spot_signature}--\ref{Sub:Analysis of the entire data-set}, where it was defined via the uncertainty of the theoretical~$\alpha_{\rm Sun}$ value~($\sigma_{\alpha}$). This $\sigma_{\alpha}$~uncertainty is brought about by the dependence of the inflection point position on the specific realisation of emergences of magnetic features. In contrast, in this section we calculate the rotation period and its uncertainty as: 

\begin{equation}\label{eq2}
    P_{\rm rot}~\pm~\delta P=~\frac{(\overline{HFIP}~\pm~2\sigma_{HFIP})}{\alpha_{\rm Sun}},
\end{equation}

\noindent where the $\sigma_{HFIP}$~value is the standard deviation of the observed distribution of inflection point positions (blue points in Fig.~\ref{Fig5}). As well as~$\sigma_{\alpha}$, $\sigma_{HFIP}$~accounts for the uncertainty due to the randomness in emergences of magnetic features (so that it does not make sense to account for~$\sigma_{\alpha}$ in Eq.~\ref{eq2}). In addition, $\sigma_{HFIP}$~also accounts for the noise in the TSI data. Hence, we utilise here Eq.~(\ref{eq2}) and $\sigma_{HFIP}$ for estimating the uncertainty of the rotation period with the GPS method.

\begin{figure*}[!ht]
\centering 
\includegraphics[trim={0 0 0 0cm},clip,width=1.0\textwidth]{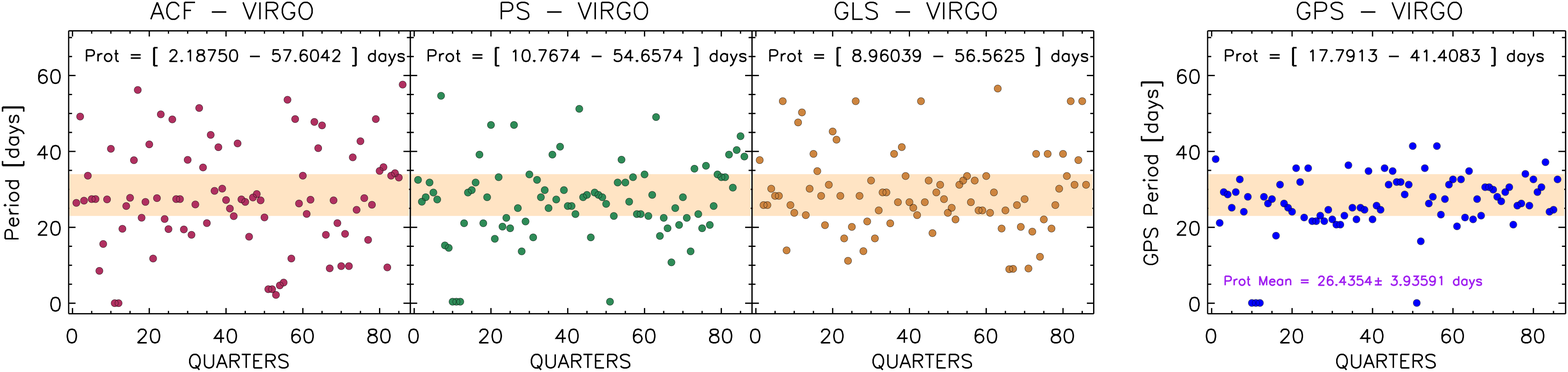}\hspace{0.0cm}
\includegraphics[trim={0 0 0 0cm},clip,width=1.0\textwidth]{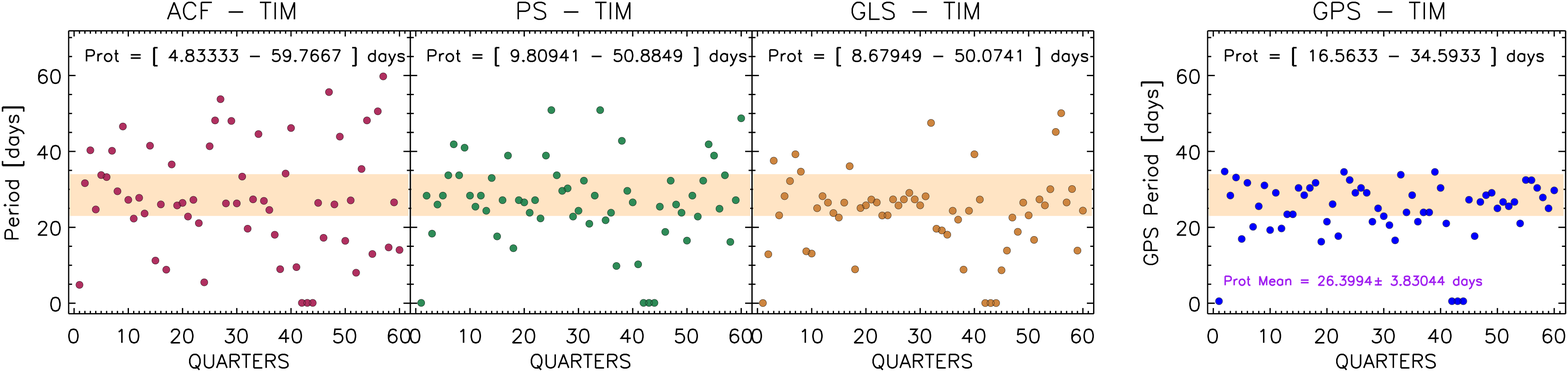}\hspace{0.0cm}
\caption{Values of the rotation periods per 90-day quarters returned by the ACF, PS, GLS, and GPS methods. The analysis is performed for the VIRGO (top panels) and TIM (bottom panels) data. Pale orange shaded areas cover the period range of [23--34]~days (see Sect.~\ref{subsec:TSIasKepler} for details). Information about the ranges of rotation period values obtained by each method for different instruments is shown near the top of each panel.}
\label{Fig6}
\end{figure*}

Figure~\ref{Fig6} compares the performance of the ACF, GLS, PS, and GPS methods. It shows one value of the rotation period per 90-day quarter determined with the four methods for 21~years of TSI by VIRGO (top panels) and 15~years of TSI by TIM (bottom panels). In addition, the pale orange bar denotes a range between 23~and 34~days, which we take to be the success range for determining the solar rotation period (which we expect would lie in the range 27--30~d within 4~d error bars). One can see that the distribution of retrieved periods is similar for data from both instruments.

The values obtained by ACF (Maroon dots) range between 2.18~days to 57.6~days for VIRGO data (see left panels of Figure~\ref{Fig6}). There are quarters where ACF can accurately detect solar rotation period, but there are many other quarters where the method fails. Overall the rotation values retrieved by the ACF lie between~23 and 34~days for 32~out of the 86~VIRGO quarters, i.e., the ACF method has a success rate of 37.2~\% when applied to the VIRGO data. For the TIM data-set the ACF obtained values are in between of 4.8~to 59~days. The success rate is 36.1~\%, as shown in Table~\ref{table:1}.

Second from the left panels in Figure~\ref{Fig6} show the performance of the PS method. The retrieved rotation periods are in between~10.7 and 54.7~days for the VIRGO data and in between 9.8~days and 50.8~days for the TIM data. The success rates for period determination are 51.2~\% and 54.1~\% for VIRGO and TIM, respectively.

Using GLS we are able to retrieve closer solar periodicities per quarter as it is shown in the right panels in Figure~\ref{Fig6}. GLS successfully retrieves the solar rotation period for 55~\% of the quarters using VIRGO data and 47~\% for TIM data. We notice lower scatter for the TIM data-set, with the returned rotation periods lying in a range between 8.6 and 50.0~days. For VIRGO data we obtain rotation period between a range of 8.9 to 56.7~days. 

In the right panels of Figure~\ref{Fig6} and Table~\ref{table:1} we show values retrieved with the GPS method. One can see that the GPS method results in less scatter for the retrieved values of the rotation period, finding rotation periods in the range [17.8-41.4]~days for VIRGO data and [16.6-34.6]~days for the TIM data-set. The GPS method achieves success rates of 88.4~\% and 82.1~\% for the VIRGO and TIM data, respectively. The regularity of the signal allow us to analyse the distribution of the high frequency inflection point and its behaviour over time.

\begin{table*}
\centering
\begin{tabular}{l c c c c c c c c c c } 
\hline
& & \\[-4pt]
D-set\textbackslash{} Meth & ACF    &   S   & GLS     & S     & PS       & S     & GPS      & S     & HFIP & LFIP \\
                           &  [d]   &  [\%] &  [d]    &  [\%] &  [d]     &  [\%] &  [d]     &  [\%] &  [d] &  [d] \\
\hline
& & \\[-3pt]
TIM SPOT     & 26.0     & --       & 27.2     & --       & 26.5      & --       & $26.6^{+2.2}_{-2.6}$ & --  & 4.21   & 19.20  \\
TIM FAC      & 26.8     & --       & 26.4     & --       & 27.1      & --       & $27.8^{+2.3}_{-2.7}$ & --  & 4.40   & 13.58  \\
VIRGO [21~Y] & --       & --       & --       & --       & --        & --       & $26.3^{+2.1}_{-2.5}$ & --  & 4.15   & --     \\
TIM [15~Y]   & --       & --       & --       & --       & --        & --       & $26.1^{+2.1}_{-2.5}$ & --  & 4.12   & --     \\
VIRGO [Q]    & 2.2-57.0 & 37.2     & 8.9-56.6 & 55.0     & 10.8-54.6 & 51.2     & 17.8-41.4  & 88.4  & 4.17$~\pm~$0.59   & --    \\
TIM [Q]      & 4.8-59.8 & 36.1     & 8.6-50.1 & 47.0     & 9.8-50.9  & 54.1     & 16.6-34.6  & 82.1  & 4.17$~\pm~$0.57   & --    \\
\hline
\end{tabular}
\vspace{0.5cm}
\caption{Compilation of the rotation period analysis for the four different methods implemented (Meth) (autocorrelation functions~(ACF), generalised Lomb-Scargle periodogram~(GLS), wavelet power spectra~(PS), and gradient of the power spectra (GPS)) and its success percentage for the different data-sets used in this work (TIM~SPOT, for the pseudo-isolated spot transit, TIM~FAC, for the pseudo facular region transit, VIRGO~[21~Y] for the entire VIRGO data-set, TIM~[15~Y] for the entire TIM~data-set, VIRGO~[Q] for the VIRGO data-set analysis per quarter, TIM~[Q], for the TIM data-set analysis per quarter).}\label{table:1} 
\end{table*}

\begin{figure*}[!ht]
\centering 
\includegraphics[trim={0 0 0 0cm},clip,width=1.0\textwidth]{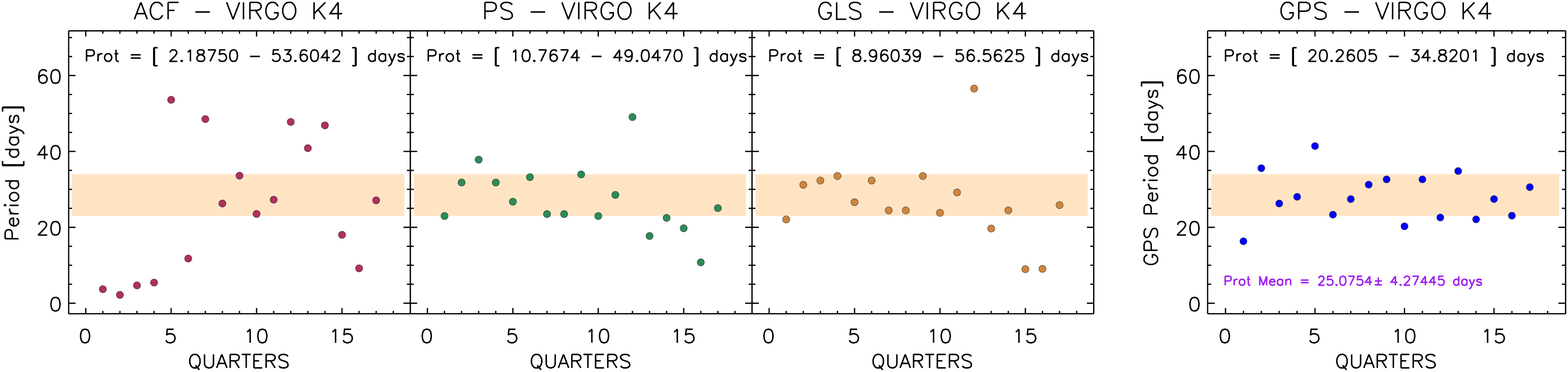}\hspace{0.0cm}
\includegraphics[trim={0 0 0 0cm},clip,width=1.0\textwidth]{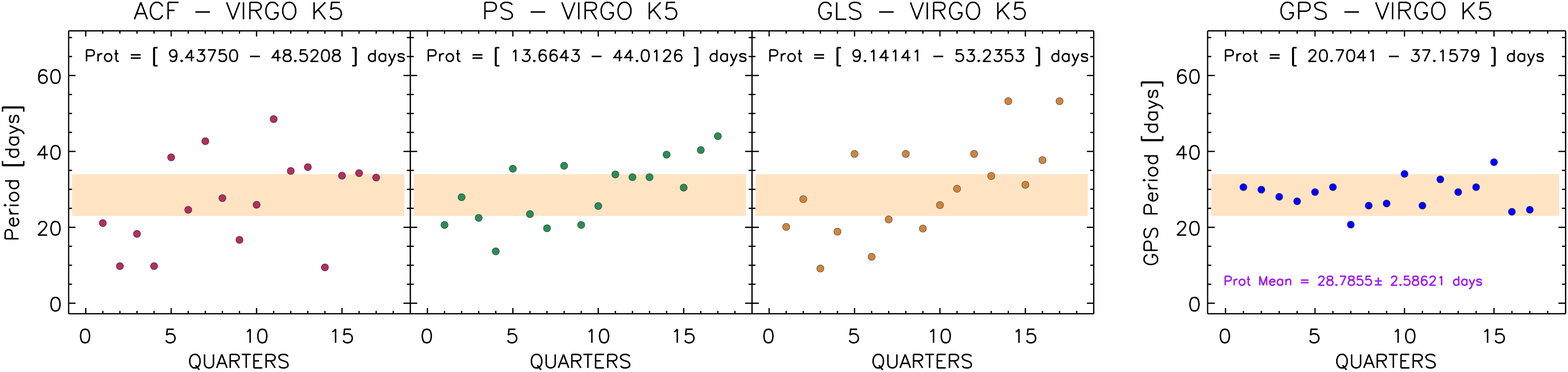}\hspace{0.0cm}
\caption{The values of the solar rotation period per 90-day quarters of the VIRGO data returned by the ACF, PS, GLS, and GPS methods (from left to right panels). Shown are the $K_4$ [2007:09:12~-~2011:07:27] and $K_5$ [2011:07:28~-~2015:06:11] Kepler-like time-span, corresponding to low and high levels of solar activity, respectively. Pale orange colour areas indicate the period range of [23--34]~days.}
\label{Fig7}
\end{figure*}

\begin{figure*}[!ht]
\centering 
\includegraphics[trim={0 0 0 0cm},clip,width=0.3\textwidth]{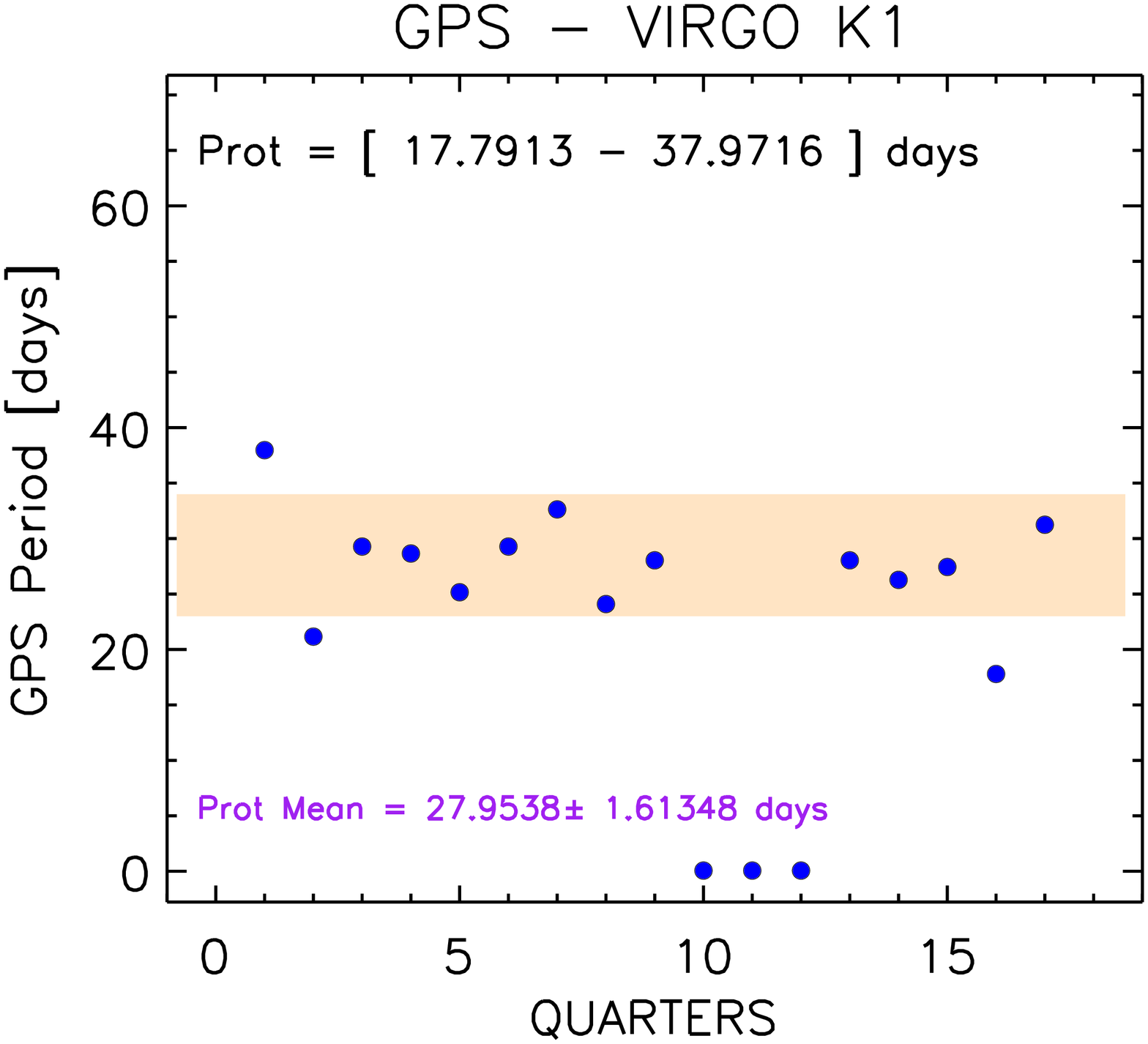}\hspace{0.0cm}
\includegraphics[trim={0 0 0 0cm},clip,width=0.3\textwidth]{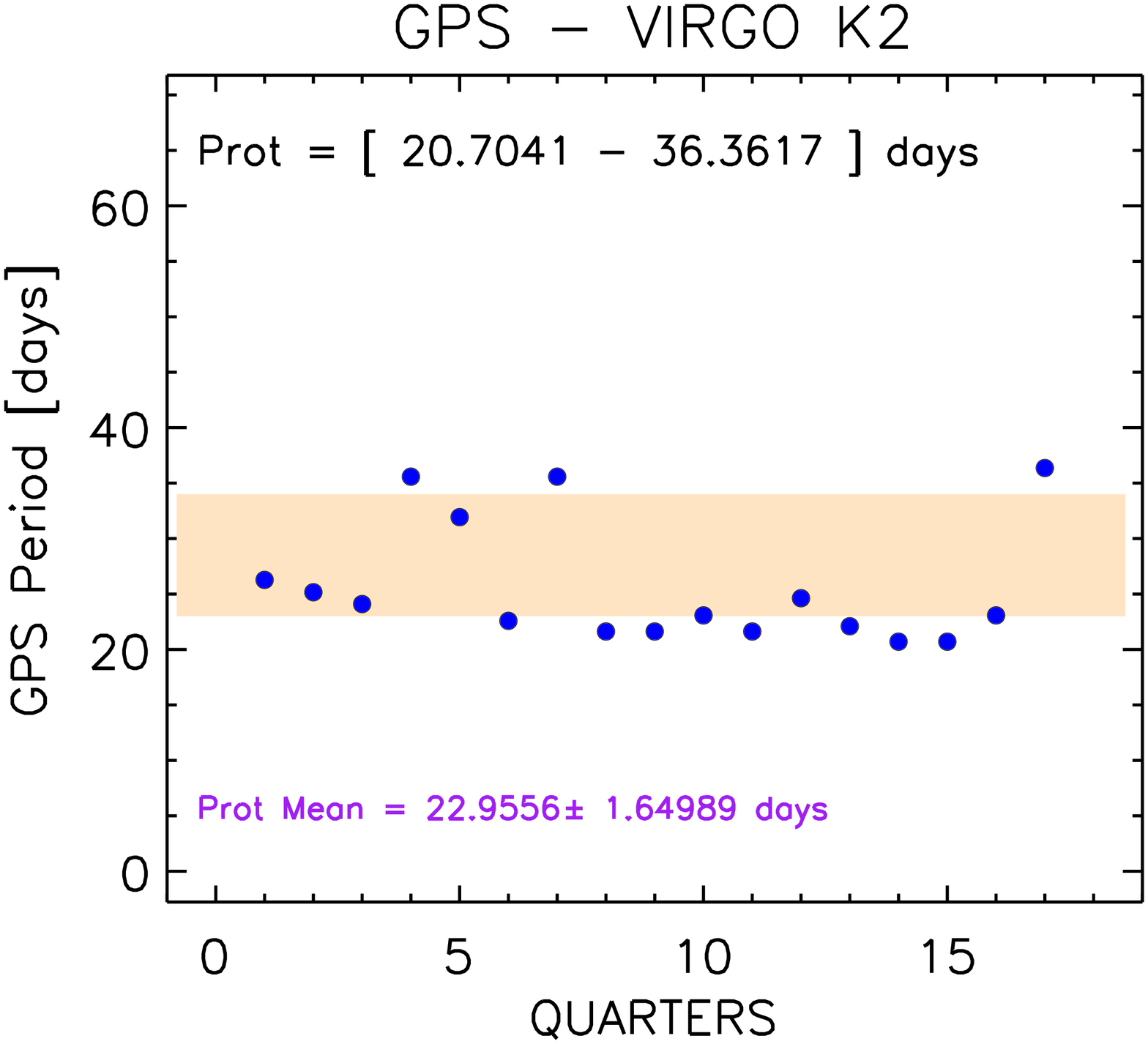}\hspace{0.0cm}\hspace{0.0cm}
\includegraphics[trim={0 0 0 1cm},clip,width=0.3\textwidth]{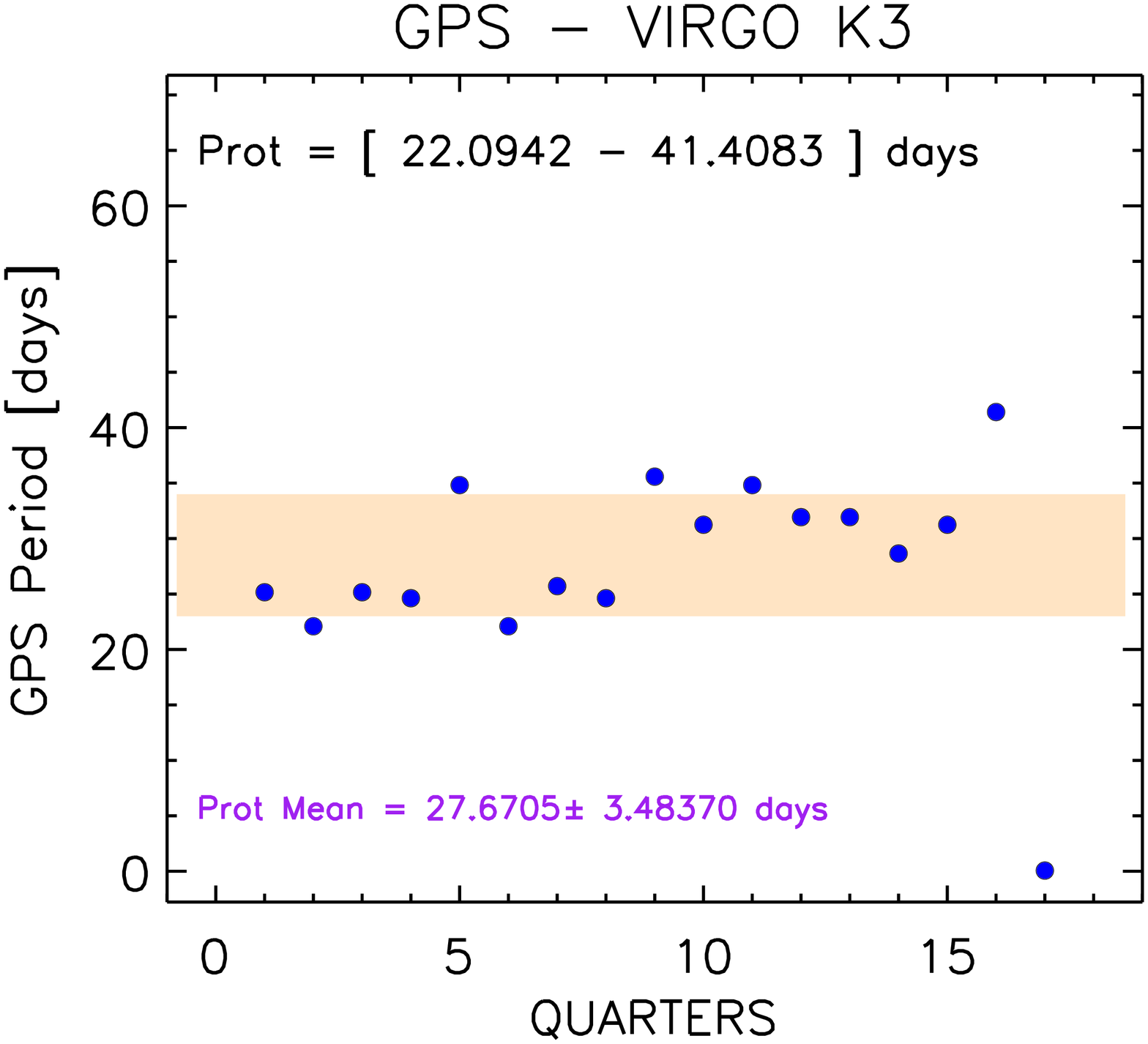}\hspace{0.0cm}\hspace{0.0cm}
\includegraphics[trim={0 0 0 0cm},clip,width=0.3\textwidth]{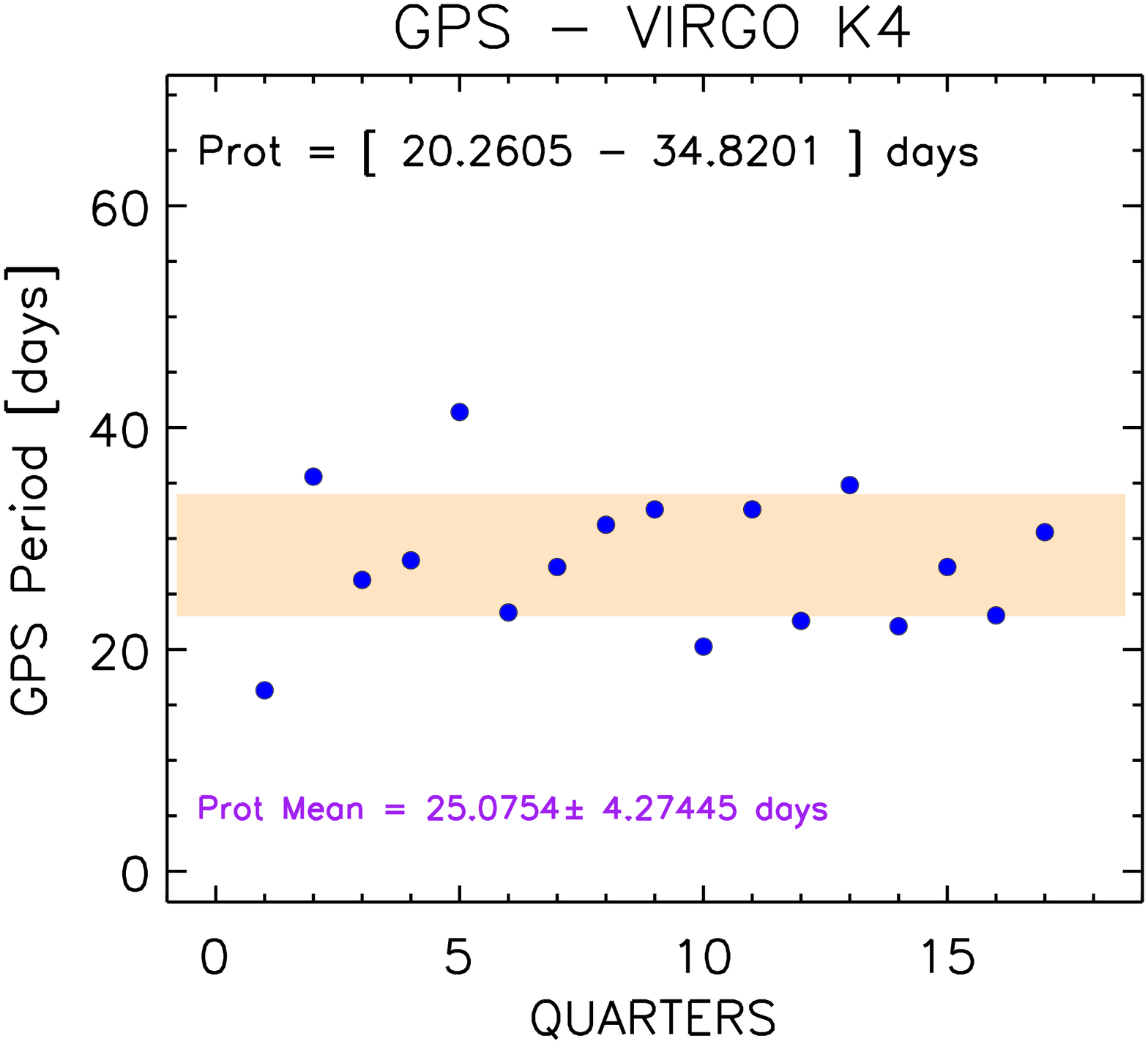}\hspace{0.0cm}\hspace{0.0cm}
\includegraphics[trim={0 0 0 0cm},clip,width=0.3\textwidth]{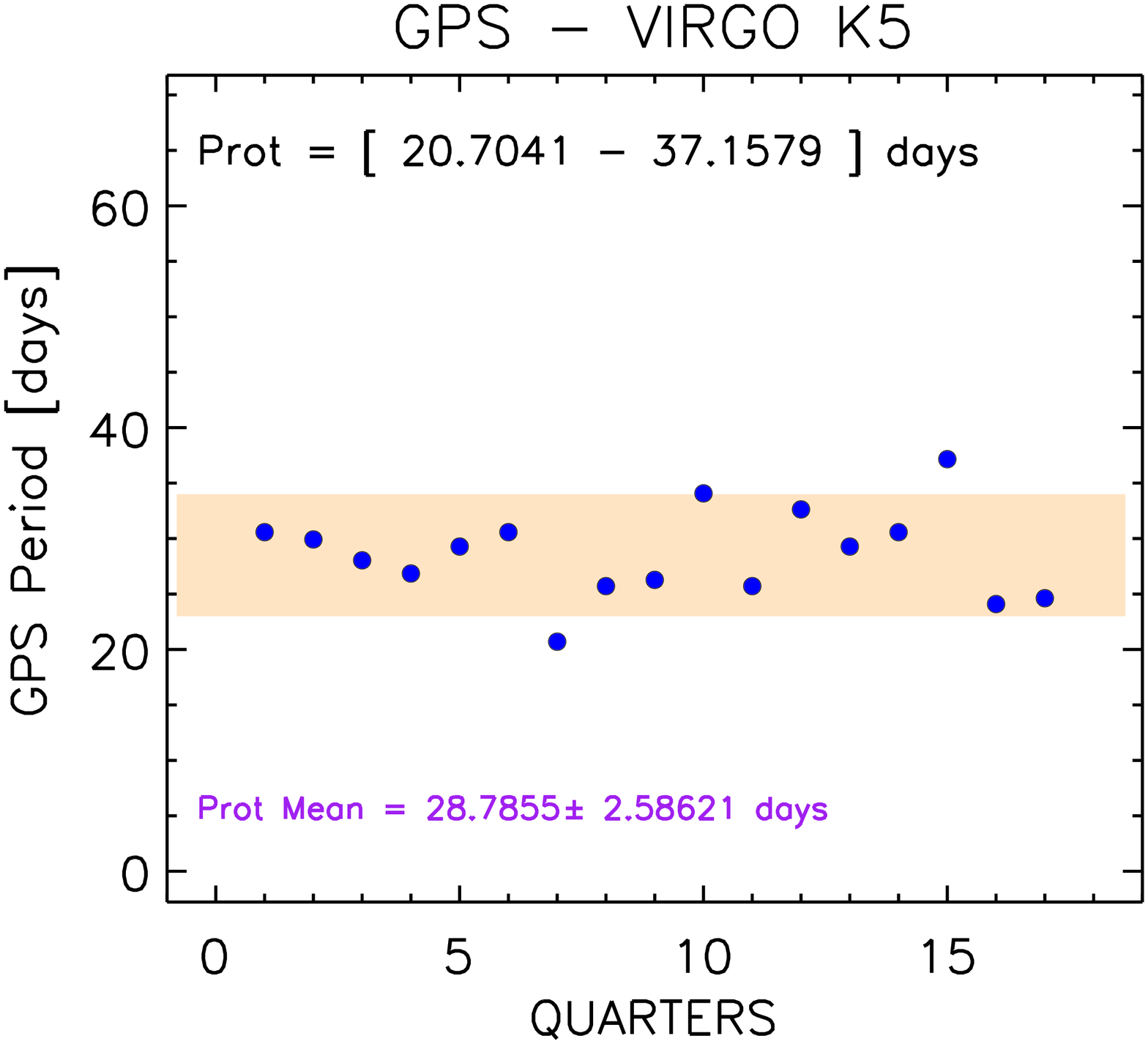}\hspace{0.0cm}\hspace{0.0cm}
\includegraphics[trim={0 0 0 0cm},clip,width=0.3\textwidth]{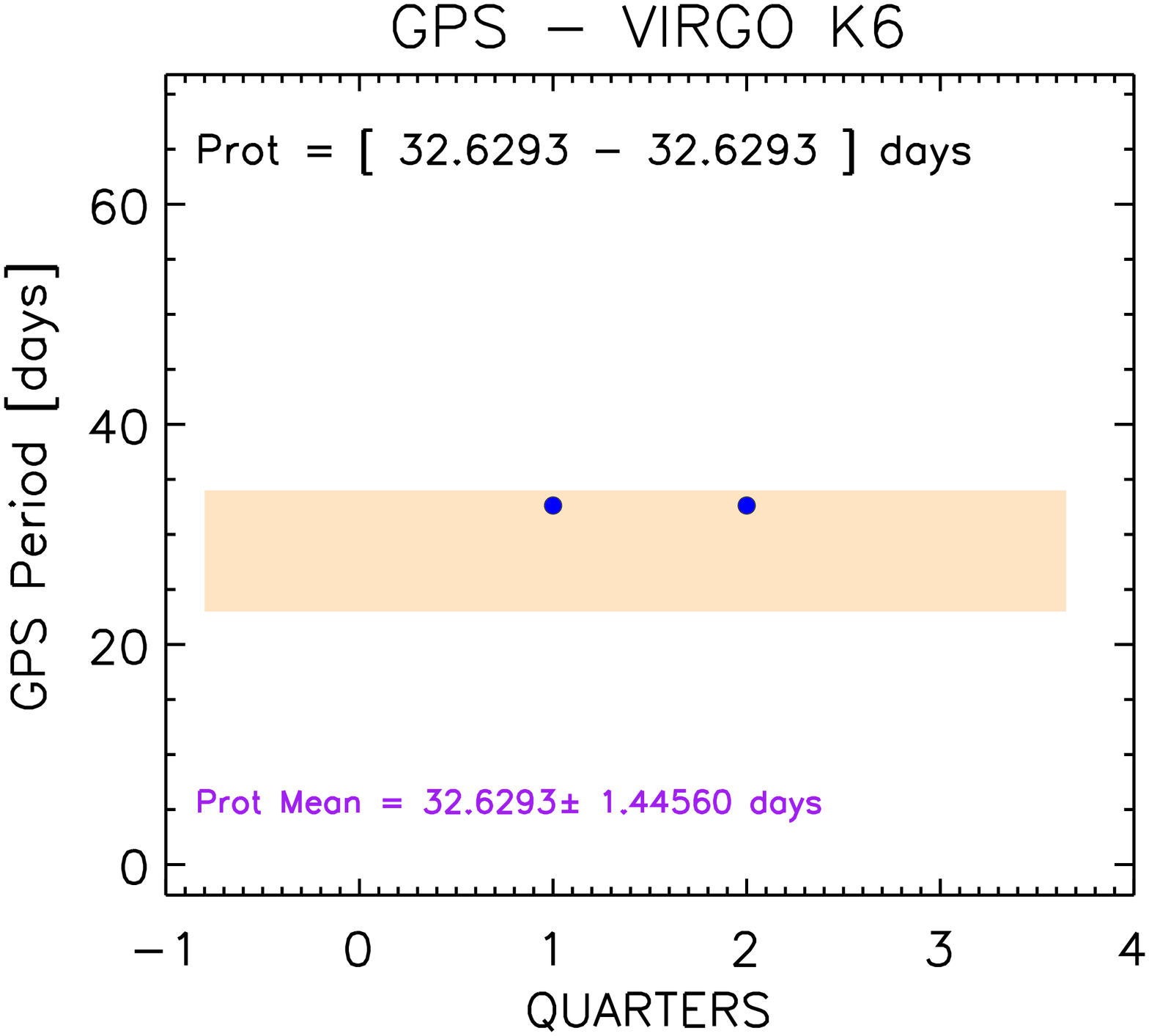}\hspace{0.0cm}\hspace{0.0cm} 
\caption{The same as Figure~\ref{Fig7} but for all Kepler-like ($K_{1}$ -- $K_{6}$) time-span of the VIRGO data. Only GPS values are shown. Blue dots represent the estimation of the rotation period per quarter obtained by GPS.}
\label{Fig8}
\end{figure*}

Regular stellar photometric observations are normally performed during unknown stellar activity stages. To characterise the detectability of the rotation period for the different activity time-spans, we test the performance of the GPS method in comparison with the ACF, GLS and PS methods for periods of relatively low and high solar activity. For this we use just VIRGO data, since it covers both periods of very high and low solar activity.

To mimic Kepler observations we split the entire period of VIRGO observations into five segments $K_{1}$~--~$K_{5}$ (the length of the segments roughly corresponds to the total duration of the 4~years of Kepler observations) and the remaining 712-day segment $K_{6}$ (see vertical dashed orange lines in Figures~\ref{Fig3},~\ref{Fig4} and, \ref{Fig5}). We then subdivided each K$_{n}$ segment in 17~quarters of 90~days each, and analyse them separately. 

The performance of all four methods is compared in Figure~\ref{Fig7} for segment $K_{4}$ (corresponding to a period of low solar activity) and $K_{5}$ (high solar activity). We observe that for the low period of activity, $K_{4}$, the values obtained per quarter for PS, GLS show less scatter than for the values shown during high levels of solar activity in $K_{5}$ segment. ACF values show similar scattered rotation values for both, high and low levels of solar activity. The GPS method recover rotation period values closer to the solar rotation period range for both $K_{4}$ and $K_{5}$ segments.

Figure~\ref{Fig8} shows the rotation period detected with the GPS method per quarter for $K_{1}$ to $K_{6}$ segments. While there is some scatter in the values of the rotation periods deduced from the analysis of segments $K_{1}$ to $K_{6}$ (in particular, values obtained in segment $K_{2}$ are 4~days lower than {\it sidereal} Carrington rotation period), Figure~\ref{Fig8} indicates that the solar rotation period can be successfully retrieved by GPS for all $K_{n}$ analysed segments.

\subsection{The impact of white noise in the inflection point position}

\begin{figure}[!ht]
\centering
\includegraphics[trim={0 0 0 0cm},clip,width=0.52\textwidth]{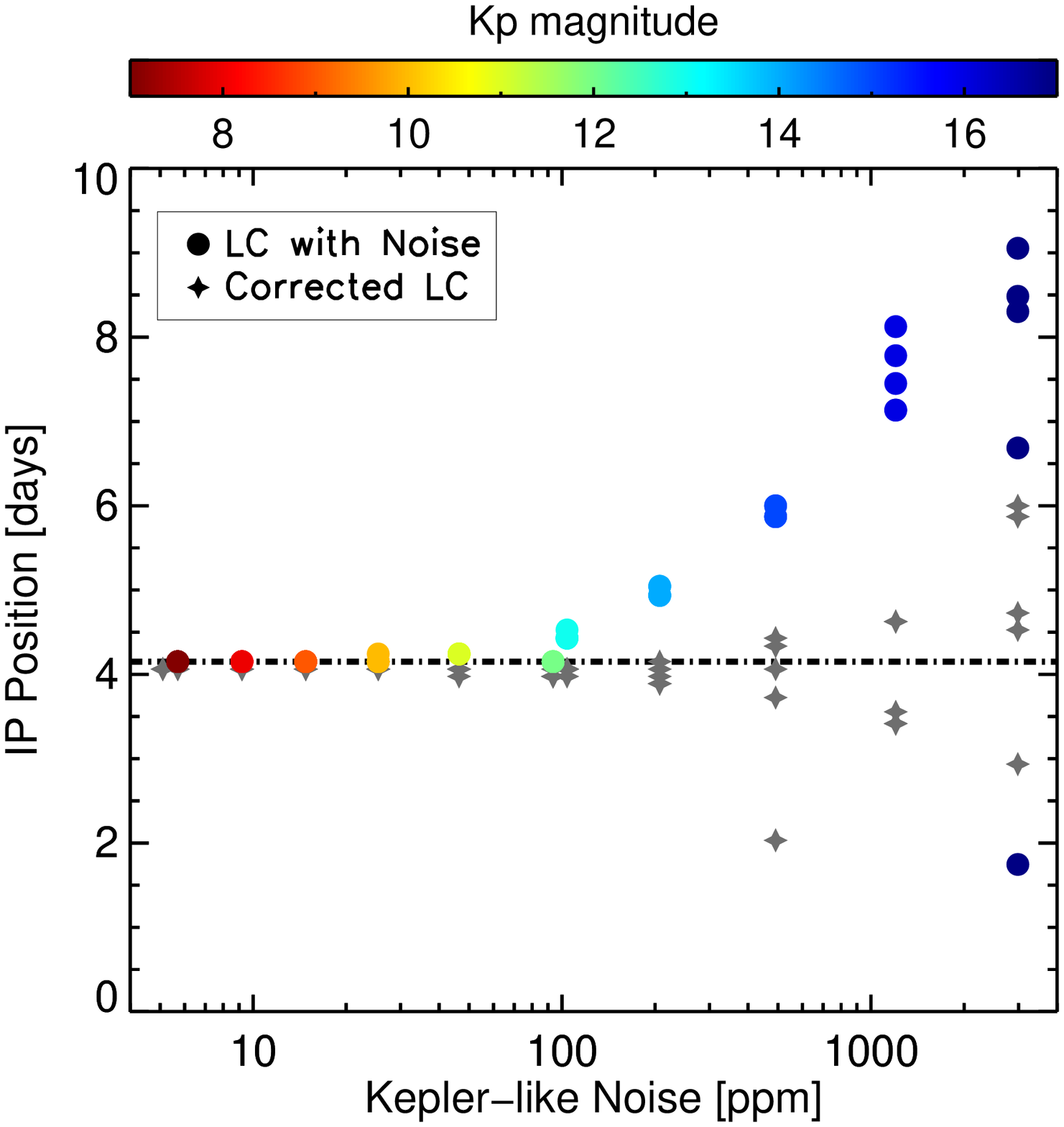}
\caption{Position of the high frequency inflection point calculated for the entire VIRGO data-set (as in Figure~\ref{Fig3}) as a function of Kepler-like white noise added to the original VIRGO data. The expected level of white noise is a function of the stellar Kepler magnitude,~Kp (see main text for more information). The colour of the dots (see colour panel in the top of the figure) indicates Kp~magnitude corresponding to the expected level of the white noise. Grey star symbols represent the position of the inflection point after the noise correction (see text for details). There are five different realisations of noise per each value of the Kp~magnitude. When the level of noise in the LC is lower than 100~ppm, around Kp=12, the values of IP values are overlapped and appear as a single point in the plot.}
\label{Fig9}
\end{figure}

In Subsection.~\ref{subsec:TSIasKepler} we have processed solar TSI data to represent the Sun as it would be observed with the time-span of Kepler observations. We have considered two TSI data-sets, one obtained by SoHO$/$VIRGO, another by TIM$/$SORCE. While the noise level in these two data-sets is rather different~\citep{Greg2016}, the positions of the inflection points are basically independent off the data-set (see, Figs.~\ref{Fig5}~and~\ref{Fig6}). This implies that our analysis is only weakly affected by the noise in TIM and VIRGO data. At the same time the noise level in Kepler data normally is significantly higher than those in the solar data.
 
Solar and stellar light-curves are recorded in a different way so that the noise sources are also substantially different. TSI is measured using radiometers while stellar photometric measurements are performed using Charged Coupled Devices (CCD). Photon detection by a CCD is a statistical process associated with several sources of noise, which can be generally approximated by Gaussian white noise.
 
To assess the impact of noise on the position of the inflection point we artificially added white noise to the VIRGO TSI data. The amplitude of the noise was chosen following the expected dependence \citep[see][]{2016ksci.rept....1V} of the noise value per specific Kepler magnitude~Kp, that is the measured source intensity observed through the Kepler bandpass. 
 
Figure~\ref{Fig9} shows the position of the high frequency inflection point as a function of the white noise level. The colours of the dots represent the Kepler magnitude~Kp. For each Kp-value (and, corresponding, amplitude of the white noise) we calculate five realisations of the noise, add it to the entire VIRGO data-set shown in Figure~\ref{Fig3}, and calculate the position of the inflection point as in Sect.~\ref{Sub:Analysis of the entire data-set}. One can see that the inflection point shifts to lower frequencies when the noise level is increased.
 
We have tested a simple method for mitigating such a shift. Namely, we utilized the fact that the power spectrum flattens at high frequencies. The power in the flattened part represents the superposition of white noise and granulation \citep{2017NatAs...1..612S}. We have calculated the mean power between periods of 1~hour and 1~day and subtracted this single value from the entire power spectrum. Then, we recalculated the position of the inflection point. These corrected positions of the inflection points are represented by grey filled star symbols  in Figure~\ref{Fig9}. One can see that our method is reasonably effective until the noise level reaches about 300-400~ppm, which corresponds to about a Kp~magnitude of~14 for 1-hour cadence light-curve. 

Beyond a level of introduced noise of 500~ppm the error in the estimation of the real high frequency inflection point location starts to become considerable, even though the location of the high frequency inflection point still gets more accurate after the correcting noise procedure, see grey star symbols in Fig.~\ref{Fig9}. For example, for a star with the level of noise expected for a Kp~magnitude of~15, the mean scatter in the high frequency inflection point value corresponds to 0.83~days, which yields to a 5.25~days deviation in the rotation period value.

\section{GPS and Skewness relation}\label{sect:Noise_and_sk}

\begin{figure*}[!ht]
\centering 
\includegraphics[trim={0 0 0 0cm},clip,width=1.0\textwidth]{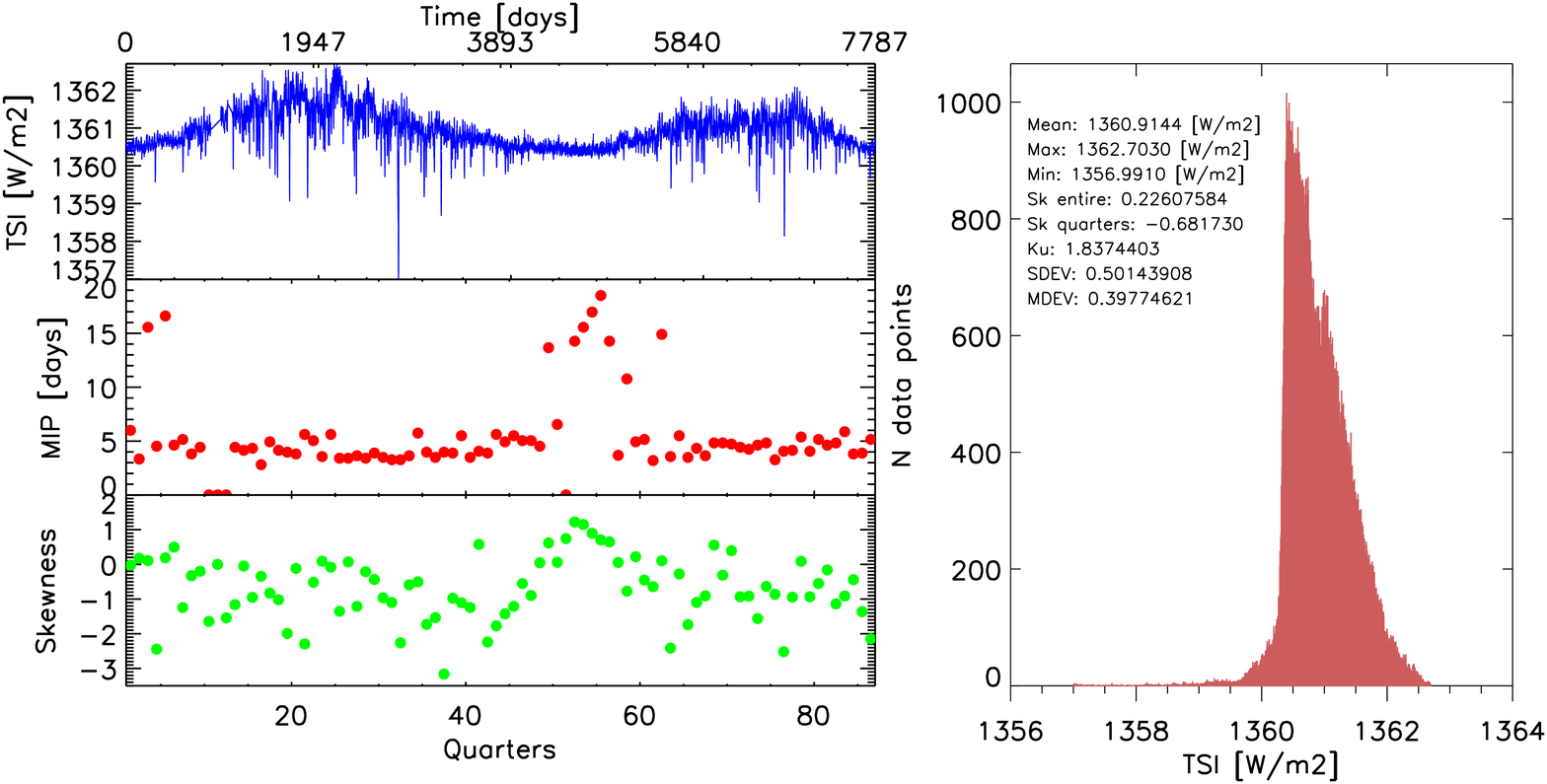}\hspace{0.0cm}
 
\caption{Top panel of left image: 21-years of TSI data gathered by VIRGO. Middle panel of left image: Positions of the maximum inflection point~(MIP) per quarter. Left-Bottom: Skewness values per quarter. Right: Distribution of the TSI values shown in the top-left panel. We list mean, maximum, and minimum TSI values, skewness for the entire data-set (Sk over the whole time-series), the mean of individual skewness values calculated per quarter (Sk~quarters, see text for more details), kurtis~(Ku), standard deviation~(SDEV), and mean deviation~(MDEV).}
\label{Fig10}
\end{figure*}

\begin{figure*}[!ht]
\centering 
\includegraphics[trim={0 0 0 0cm},clip,width=1.\textwidth]{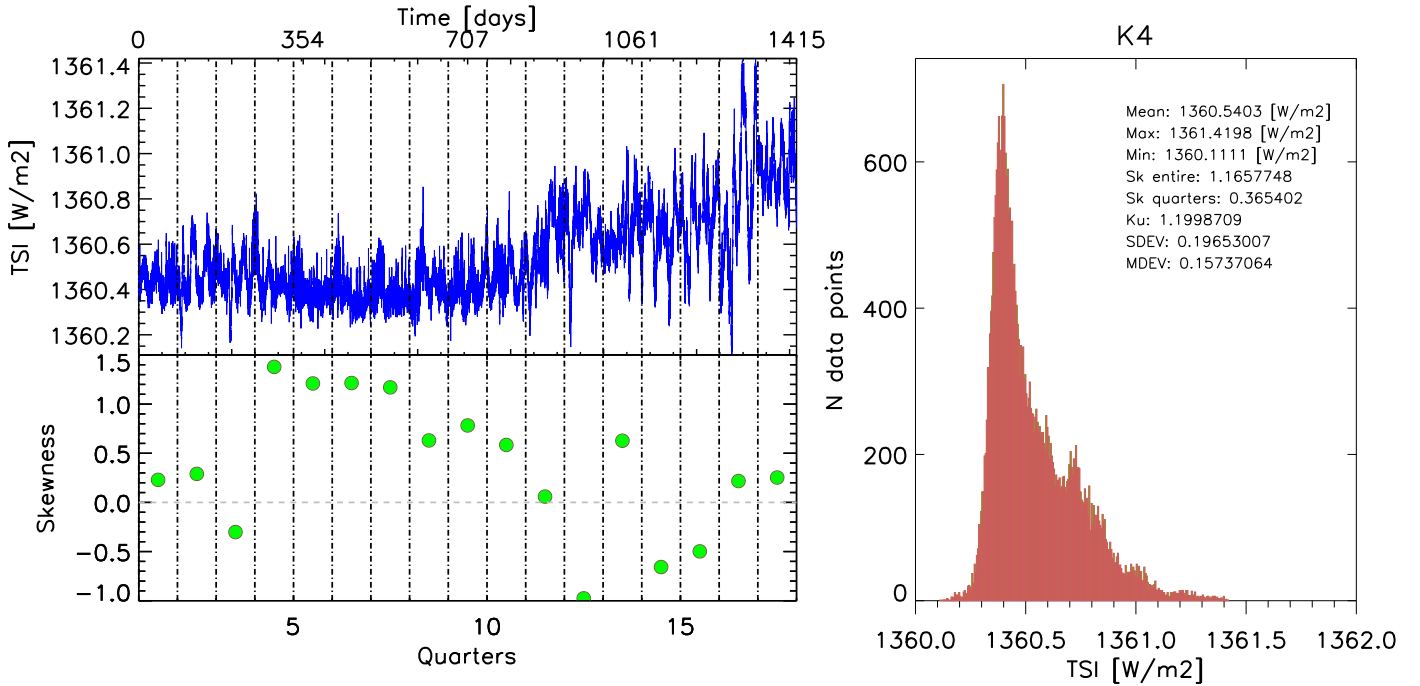}\hspace{0.0cm}
 
\caption{Skewness analysis for low-activity Kepler-like time-span $K_{4}$. Upper left panel: VIRGO TSI time-series during the $K_{4}$~season, which consists of 1415~days emulating 17~Kepler quarters. Lower left panel: Skewness values per quarter. Right: the same as right panel in Figure~\ref{Fig10} but showing only TSI values in $K_4$ season.}
\label{Fig11}
\end{figure*}

\begin{figure*}[!ht]
\centering 
 
\includegraphics[trim={0 0 0 0cm},clip,width=1.\textwidth]{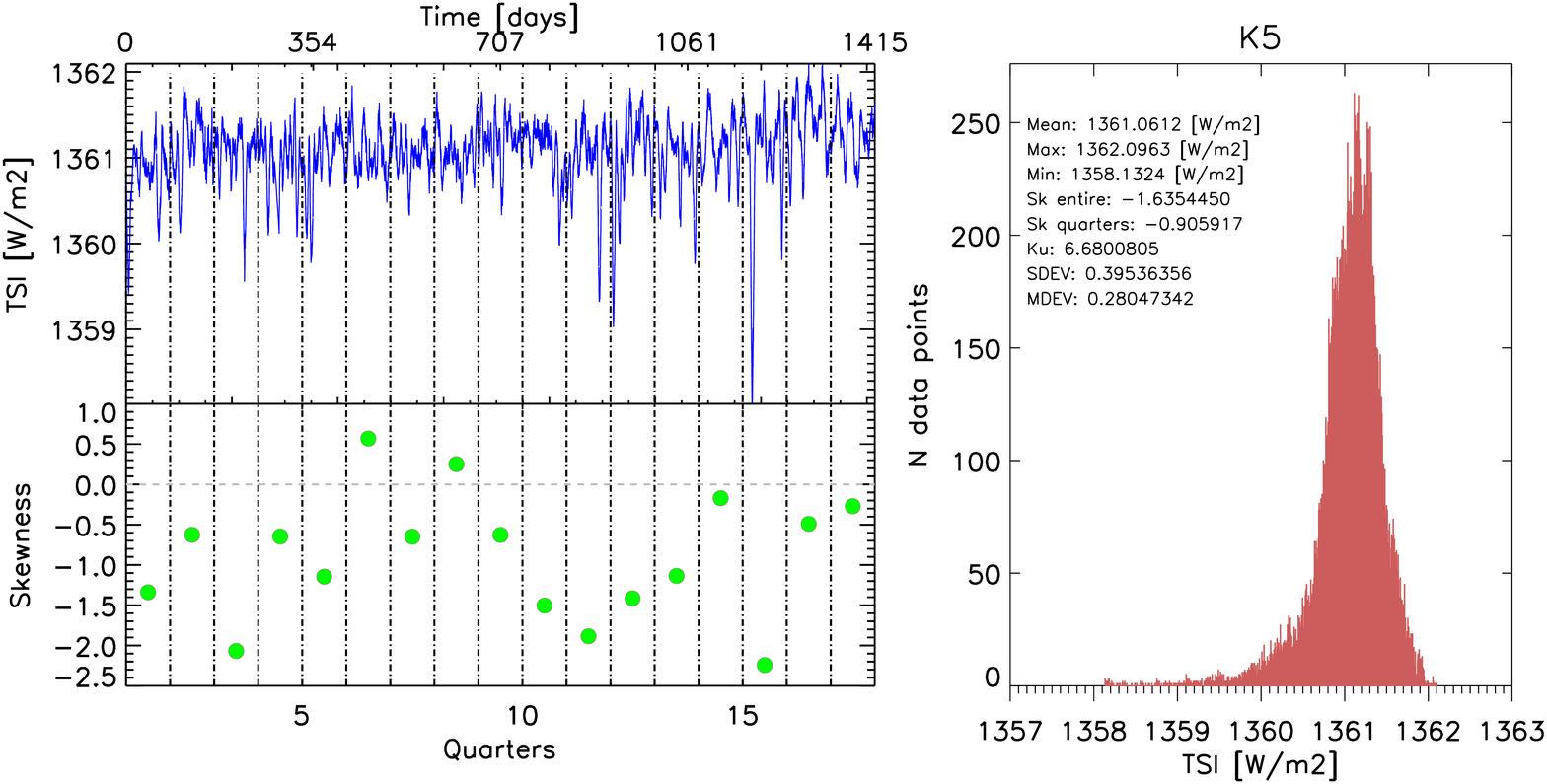}\hspace{0.0cm}
\caption{The same as Figure~\ref{Fig11} but for high-activity Kepler-like time-span $K_{5}$.}
\label{Fig12}
\end{figure*}
 
Distinguishing between facular- and spot-dominated regimes of brightness variability is important for understanding the structure of the stellar magnetic field and for identifying biases in determination of stellar rotation periods. 

While solar rotation variability is predominantly spot-dominated, there are also periods of facular domination (see Sect.~\ref{subsec:fac_signature}). In Sects.~\ref{subsec:Spot_signature}~and~\ref{subsec:fac_signature} we demonstrated that the GPS spectrum has a different profile depending on whether variability is facular- or spot-dominated. In this section we show that in the solar case these two regimes can also be distinguished based on the skewness of the distribution of TSI values. This suggests that skewness can be a good indicator of the variability regime for low-activity stars like the Sun. 

For a data-set of TSI values the skewness can give us valuable information about the distribution of maximum and minimum values. In other words when we have a decrease of the intensity due spots the distribution will be skewed preferentially towards the left side (i.e., to the lower values) of the maximum peak of the distribution. When an increase of intensity is registered in the light-curve due the presence of brighter facular regions the skewness will shift to the right side (i.e., to the higher values) of the distribution. The skewness of a distribution can tell us about its degree of symmetry.

In order to analyse the relation between skewness and the regime of solar variability we calculate skewness of the TSI values in each of the 90-day quarters introduced in Subsection.~\ref{subsec:TSIasKepler}. In the upper left panel of Figure~\ref{Fig10} we show the 21-year span of VIRGO TSI data. In the middle left panel we illustrate the location of the maximum inflection point (see Sect.~\ref{Sub:Analysis of the entire data-set}) per quarter. As discussed in Sects.~\ref{subsec:Spot_signature}~and~\ref{subsec:fac_signature}, the maximum inflection point (MIP) corresponds to the low frequency inflection point for faculae-dominated regimes and high frequency inflection point for spot-dominated regimes. The bottom left panel shows the skewness values for all 90-day quarters. One can see that periods of low solar activity, when TSI variability is mainly brought about by faculae \citep[see, e.g., discussion in][]{Shapiroetal2016}, simultaneously correspond to positive skewness and maximum GPS value reached at the low frequency inflection point. The observed scatter in skewness values is higher scatter than the scatter in the MIP by GPS. This implies that the inflection points analysis provides a better indication of when the LC is mainly drawn by spot or facular components.

In the right panel of Figure~\ref{Fig10} we show the distribution of TSI values for the entire VIRGO data. The skewness for the entire data-set, $Sk_{E}=0.23$, is positive. This is because the skewness value of the entire data-set is affected by the TSI variability on the timescale of the 11-year cycle, which is faculae-dominated. To remove the contribution from the 11-year variability we can also calculate skewness by averaging all individual values per 90-day quarters. Since rotation TSI variability is mainly spot-dominated we then get a negative value of $Sk_{Q}=-0.69$.

Figures~\ref{Fig11} and \ref{Fig12} show the skewness analysis for minimum and maximum activity segments, $K_{4}$ and $K_{5}$ , respectively. The comparison between the light-curve segmented in quarters and its respective skewness values are in agreement with conclusions drawn analysing the entire VIRGO data-set. In particular, one can see that quarters with prominent positive excursions of brightness caused by faculae correspond to positive skewness, while quarters with negative excursions caused by spot correspond to negative values of skewness. Clearly, also the skewness of brightness distribution in the entire $K_{4}$ segment (corresponding to the minimum of solar activity) is positive, while skewness values for the $K_{5}$ segment with higher value of activity is negative. We will extend the combined, skewness and GPS, analysis to stars observed by Kepler and TESS in the forthcoming publications.
 
\section{Discussion \& Summary}\label{sect:summary}

\noindent The determination of rotation periods of stars with activity levels similar to that of our Sun is a challenging task, even when using high quality data from space-borne photometric missions. In \cite{paperI} we have proposed the GPS method specifically aimed at the determination of periods in old inactive stars, like our Sun. The main idea of the method is to calculate the gradient of the power spectrum of stellar brightness variations and identify the inflection point, i.e., the point where concavity of the power spectrum changes its sign. The stellar rotation period can then be determined by applying a scaling coefficient to the position of the inflection point.

We have applied the GPS method to the available measured records of solar brightness (specifically the total solar irradiance) and compared its performance to that of other methods routinely utilized for the determination of stellar rotation periods.

There are time intervals when solar light-curve has a regular pattern, the GPS and other methods, return correct value of solar rotation period. These intervals correspond to low values of solar activity when variability is either brought about by long-living faculae or nested sunspots. However, most of the time, solar brightness variations are attributed to superposition of simultaneous contributions from several bright and dark magnetic features with random phases. We have shown that this leads to a failure of other methods to identify a clear signal of the rotation period. At the same time, the GPS method still allows an accurate determination of the rotation period of the Sun independently of its activity level and the number of features contributing to brightness variability and of the ratio of facular to sunspot area. 

In particular, we have shown that GPS method returns accurate values of solar rotation period for most of the time-span of SoHO/VIRGO and SORCE/TIM measurements, with exception of several intervals affected by the absence of data. We found that when the entire 21-year VIRGO and 15-year TIM data-sets are split in Kepler-like 90-day quarters and inflection points are calculated for each of the quarters, the maximum of the distribution of the inflection point positions peaks at $4.17~\pm~0.59$~days for VIRGO data-set and $4.17~\pm~0.57$~days for TIM data-set (see Figure~\ref{Fig5}). This results in a determination of the solar rotation period of $26.4~\pm~3.7$~days and $26.4~\pm~3.6$~days for VIRGO and TIM data-sets respectively. In a series of typical Kepler-like observations of the Sun, the GPS method can correctly determine the rotation period in more than 80~\% of the cases while this value is about 50~\% for GLS and below 40~\% for ACF.

Typically solar variability on timescales up to a few months is spot-dominated. However, there are also time intervals when it is faculae-dominated (see, e.g.,  Figure~\ref{Fig2}). We have shown that these regimes can be distinguished from the GPS profile thanks to substantially different centre-to-limb variations of facular and spot contrasts. Furthermore, the two regimes can be separated by analysing the comparison between the inflection point location from GPS and the skewness of light-curves: the bright faculae lead to positively skewed light-curves and a stronger signal at the low frequency inflection point, while dark spots lead to negatively skewed light-curves and a dominant signal at the low frequency inflection point. However, the skewness values in Figure~\ref{Fig10} show higher scatter than the IP by GPS. This implies that the IP by GPS provide a better indication of when the LC is mainly drawn by spot or facular components.

We construe the success of the GPS method in the solar case as an indication that it can be applied to reliably determine rotation periods in low-activity stars like the Sun, where other methods generally fail. Furthermore, our analysis demonstrates that photometric records alone can be used to identify the regime of stellar variability, i.e., whether it is dominated by the effects of spots or of faculae. In subsequent papers we will apply GPS method to determine rotation periods and regimes of the variability of Kepler and long term follow up of TESS stars.

\begin{acknowledgements}
We would like to thank the referee for the constructive comments which helped to improve the quality of this paper. The analysis presented in this Paper~Is based on new scale version~6.4 observations collected by the VIRGO Experiment on the cooperative ESA/NASA Mission~SoHO, provided by the VIRGO team through PMOD/WRC, Davos, Switzerland. In addition orbital-averaged version~17, level~3.0 data from the Total Irradiance Monitor~(TIM) on the NASA Earth Observing System (EOS) SOlar Radiation and Climate Experiment~(SORCE) where analysed. This work was supported by the International Max-Planck Research School~(IMPRS) for Solar System Science at the University of G{\"o}ttingen and European Research Council under the European Union Horizon~2020 research and innovation program (grant agreement by the No. 715947). Financial support was also provided by the Brain Korea 21 plus program through the National Research Foundation funded by the Ministry of Education of Korea and by the German Federal Ministry of Education and Research under project 01LG1209A. We would like to thank the International Space Science Institute, Bern, for their support of science team~446 and the resulting helpful discussions.
 
\end{acknowledgements}

\bibliographystyle{aasjournal}
\bibliography{Biblio}

\begin{thebibliography}{}
\expandafter\ifx\csname natexlab\endcsname\relax\def\natexlab#1{#1}\fi
\providecommand{\url}[1]{\href{#1}{#1}}

\bibitem[{{Aigrain} {et~al.}(2015){Aigrain}, {Llama}, {Ceillier}, {Chagas},
  {Davenport}, {Garc{\'{\i}}a}, {Hay}, {Lanza}, {McQuillan}, {Mazeh}, {de
  Medeiros}, {Nielsen}, \& {Reinhold}}]{2015MNRAS.450.3211A}
{Aigrain}, S., {Llama}, J., {Ceillier}, T., {et~al.} 2015, \mnras, 450, 3211

\bibitem[{{Angus} {et~al.}(2018){Angus}, {Morton}, {Aigrain}, {Foreman-Mackey},
  \& {Rajpaul}}]{2018MNRAS.474.2094A}
{Angus}, R., {Morton}, T., {Aigrain}, S., {Foreman-Mackey}, D., \& {Rajpaul},
  V. 2018, \mnras, 474, 2094

\bibitem[{{Barnes}(2003)}]{2003ApJ...586..464B}
{Barnes}, S.~A. 2003, \apj, 586, 464

\bibitem[{{Basri} {et~al.}(2013){Basri}, {Walkowicz}, \&
  {Reiners}}]{basrietal2013}
{Basri}, G., {Walkowicz}, L.~M., \& {Reiners}, A. 2013, \apj, 769, 37

\bibitem[{{Bord{\'e}} {et~al.}(2003){Bord{\'e}}, {Rouan}, \&
  {L{\'e}ger}}]{2003A&A...405.1137B}
{Bord{\'e}}, P., {Rouan}, D., \& {L{\'e}ger}, A. 2003, \aap, 405, 1137

\bibitem[{{Borucki} {et~al.}(2010){Borucki}, {Koch}, {Basri}, {Batalha},
  {Brown}, {Caldwell}, {Caldwell}, {Christensen-Dalsgaard}, {Cochran},
  {DeVore}, {Dunham}, {Dupree}, {Gautier}, {Geary}, {Gilliland}, {Gould},
  {Howell}, {Jenkins}, {Kondo}, {Latham}, {Marcy}, {Meibom}, {Kjeldsen},
  {Lissauer}, {Monet}, {Morrison}, {Sasselov}, {Tarter}, {Boss}, {Brownlee},
  {Owen}, {Buzasi}, {Charbonneau}, {Doyle}, {Fortney}, {Ford}, {Holman},
  {Seager}, {Steffen}, {Welsh}, {Rowe}, {Anderson}, {Buchhave}, {Ciardi},
  {Walkowicz}, {Sherry}, {Horch}, {Isaacson}, {Everett}, {Fischer}, {Torres},
  {Johnson}, {Endl}, {MacQueen}, {Bryson}, {Dotson}, {Haas}, {Kolodziejczak},
  {Van Cleve}, {Chandrasekaran}, {Twicken}, {Quintana}, {Clarke}, {Allen},
  {Li}, {Wu}, {Tenenbaum}, {Verner}, {Bruhweiler}, {Barnes}, \&
  {Prsa}}]{2010Sci...327..977B}
{Borucki}, W.~J., {Koch}, D., {Basri}, G., {et~al.} 2010, Science, 327, 977

\bibitem[{{Brouwer} \& {Zwaan}(1990)}]{1990SoPh..129..221B}
{Brouwer}, M.~P., \& {Zwaan}, C. 1990, \solphys, 129, 221

\bibitem[{{Charbonneau}(2010)}]{2010LRSP....7....3C}
{Charbonneau}, P. 2010, Living Reviews in Solar Physics, 7, 3

\bibitem[{{Davenport}(2017)}]{Davenport2017}
{Davenport}, J.~R.~A. 2017, \apj, 835, 16

\bibitem[{{Dumusque} {et~al.}(2011){Dumusque}, {Udry}, {Lovis}, {Santos}, \&
  {Monteiro}}]{2011A&A...525A.140D}
{Dumusque}, X., {Udry}, S., {Lovis}, C., {Santos}, N.~C., \& {Monteiro},
  M.~J.~P.~F.~G. 2011, \aap, 525, A140

\bibitem[{Ermolli {et~al.}(2013)Ermolli, Matthes, Dudok~de Wit, Krivova,
  Tourpali, Weber, Unruh, Gray, Langematz, Pilewskie, Rozanov, Schmutz,
  Shapiro, Solanki, \& Woods}]{acp-13-3945-2013}
Ermolli, I., Matthes, K., Dudok~de Wit, T., {et~al.} 2013, Atmospheric
  Chemistry and Physics, 13, 3945.
\newblock \url{https://www.atmos-chem-phys.net/13/3945/2013/}

\bibitem[{{Farge}(1992)}]{1992AnRFM..24..395F}
{Farge}, M. 1992, Annual Review of Fluid Mechanics, 24, 395

\bibitem[{{Faria} {et~al.}(2019){Faria}, {Adibekyan}, {Amazo-G{\'o}mez},
  {Barros}, {Camacho}, {Demangeon}, {Figueira}, {Mortier}, {Oshagh}, {Pepe},
  {Santos}, {Gomes da Silva}, {Costa Silva}, {Sousa}, {Ulmer-Moll}, \&
  {Viana}}]{2019arXiv191111714F}
{Faria}, J.~P., {Adibekyan}, V., {Amazo-G{\'o}mez}, E.~M., {et~al.} 2019, arXiv
  e-prints, arXiv:1911.11714

\bibitem[{{Fligge} {et~al.}(2000){Fligge}, {Solanki}, \&
  {Unruh}}]{2000A&A...353..380F}
{Fligge}, M., {Solanki}, S.~K., \& {Unruh}, Y.~C. 2000, \aap, 353, 380

\bibitem[{{Foreman-Mackey} {et~al.}(2017{\natexlab{a}}){Foreman-Mackey},
  {Agol}, {Ambikasaran}, \& {Angus}}]{2017AJ....154..220F}
{Foreman-Mackey}, D., {Agol}, E., {Ambikasaran}, S., \& {Angus}, R.
  2017{\natexlab{a}}, \aj, 154, 220

\bibitem[{{Foreman-Mackey} {et~al.}(2017{\natexlab{b}}){Foreman-Mackey},
  {Agol}, {Angus}, {Brewer}, {Austin}, {Meierjurgen Farr}, {Guillochon},
  {Czekala}, \& {Casey}}]{2017zndo...1048287F}
{Foreman-Mackey}, D., {Agol}, E., {Angus}, R., {et~al.} 2017{\natexlab{b}},
  {Dfm/Celerite: Celerite V0.3.0}, vv0.3.0,  Zenodo, doi:10.5281/zenodo.1048287

\bibitem[{{Fr{\"o}hlich} {et~al.}(1997){Fr{\"o}hlich}, {Crommelynck}, {Wehrli},
  {Anklin}, {Dewitte}, {Fichot}, {Finsterle}, {Jim{\'e}nez}, {Chevalier}, \&
  {Roth}}]{1997SoPh..175..267F}
{Fr{\"o}hlich}, C., {Crommelynck}, D.~A., {Wehrli}, C., {et~al.} 1997,
  \solphys, 175, 267

\bibitem[{{Gaizauskas} {et~al.}(1994){Gaizauskas}, {Harvey}, \&
  {Proulx}}]{1994ApJ...422..883G}
{Gaizauskas}, V., {Harvey}, K.~L., \& {Proulx}, M. 1994, \apj, 422, 883

\bibitem[{{Garc{\'{\i}}a} {et~al.}(2009){Garc{\'{\i}}a}, {R{\'e}gulo},
  {Samadi}, {Ballot}, {Barban}, {Benomar}, {Chaplin}, {Gaulme}, {Appourchaux},
  {Mathur}, {Mosser}, {Toutain}, {Verner}, {Auvergne}, {Baglin}, {Baudin},
  {Boumier}, {Bruntt}, {Catala}, {Deheuvels}, {Elsworth}, {Jim{\'e}nez-Reyes},
  {Michel}, {P{\'e}rez Hern{\'a}ndez}, {Roxburgh}, \&
  {Salabert}}]{2009A&A...506...41G}
{Garc{\'{\i}}a}, R.~A., {R{\'e}gulo}, C., {Samadi}, R., {et~al.} 2009, \aap,
  506, 41

\bibitem[{{Garc{\'{\i}}a} {et~al.}(2014){Garc{\'{\i}}a}, {Ceillier},
  {Salabert}, {Mathur}, {van Saders}, {Pinsonneault}, {Ballot}, {Beck},
  {Bloemen}, {Campante}, {Davies}, {do Nascimento}, {Mathis}, {Metcalfe},
  {Nielsen}, {Su{\'a}rez}, {Chaplin}, {Jim{\'e}nez}, \& {Karoff}}]{Garcia2014}
{Garc{\'{\i}}a}, R.~A., {Ceillier}, T., {Salabert}, D., {et~al.} 2014, \aap,
  572, A34

\bibitem[{{He} {et~al.}(2018){He}, {Wang}, {Yan}, \&
  {Yun}}]{2018IAUS..335....7H}
{He}, H., {Wang}, H., {Yan}, Y., \& {Yun}, D. 2018, in IAU Symposium, Vol. 335,
  Space Weather of the Heliosphere: Processes and Forecasts, ed. C.~{Foullon}
  \& O.~E. {Malandraki}, 7--10

\bibitem[{{He} {et~al.}(2015){He}, {Wang}, \& {Yun}}]{2015ApJS..221...18H}
{He}, H., {Wang}, H., \& {Yun}, D. 2015, \apjs, 221, 18

\bibitem[{{I{\c s}{\i}k} {et~al.}(2018){I{\c s}{\i}k}, {Solanki}, {Krivova}, \&
  {Shapiro}}]{2018arXiv181006728I}
{I{\c s}{\i}k}, E., {Solanki}, S.~K., {Krivova}, N.~A., \& {Shapiro}, A.~I.
  2018, ArXiv e-prints, arXiv:1810.06728

\bibitem[{{Kopp}(2014)}]{2014JSWSC...4A..14K}
{Kopp}, G. 2014, Journal of Space Weather and Space Climate, 4, A14

\bibitem[{{Kopp}(2016)}]{Greg2016}
---. 2016, Journal of Space Weather and Space Climate, 6, A30

\bibitem[{{Kopp} {et~al.}(2005{\natexlab{a}}){Kopp}, {Heuerman}, \&
  {Lawrence}}]{2005SoPh..230..111K}
{Kopp}, G., {Heuerman}, K., \& {Lawrence}, G. 2005{\natexlab{a}}, \solphys,
  230, 111

\bibitem[{{Kopp} \& {Lawrence}(2005)}]{2005SoPh..230...91K}
{Kopp}, G., \& {Lawrence}, G. 2005, \solphys, 230, 91

\bibitem[{{Kopp} {et~al.}(2005{\natexlab{b}}){Kopp}, {Lawrence}, \&
  {Rottman}}]{2005SoPh..230..129K}
{Kopp}, G., {Lawrence}, G., \& {Rottman}, G. 2005{\natexlab{b}}, \solphys, 230,
  129

\bibitem[{{Krivova} {et~al.}(2011){Krivova}, {Solanki}, \&
  {Schmutz}}]{2011A&A...529A..81K}
{Krivova}, N.~A., {Solanki}, S.~K., \& {Schmutz}, W. 2011, \aap, 529, A81

\bibitem[{{Lomb}(1976)}]{1976Ap&SS..39..447L}
{Lomb}, N.~R. 1976, \apss, 39, 447

\bibitem[{{Mart{\'{\i}}nez Pillet} {et~al.}(1993){Mart{\'{\i}}nez Pillet},
  {Moreno-Insertis}, \& {Vazquez}}]{decay2}
{Mart{\'{\i}}nez Pillet}, V., {Moreno-Insertis}, F., \& {Vazquez}, M. 1993,
  \aap, 274, 521

\bibitem[{{McQuillan} {et~al.}(2013){McQuillan}, {Aigrain}, \&
  {Mazeh}}]{McQuillan2013a}
{McQuillan}, A., {Aigrain}, S., \& {Mazeh}, T. 2013, \mnras, 432, 1203

\bibitem[{{McQuillan} {et~al.}(2014){McQuillan}, {Mazeh}, \&
  {Aigrain}}]{2014ApJS..211...24M}
{McQuillan}, A., {Mazeh}, T., \& {Aigrain}, S. 2014, \apjs, 211, 24

\bibitem[{{Newton} {et~al.}(2017){Newton}, {Irwin}, {Charbonneau}, {Berlind},
  {Calkins}, \& {Mink}}]{2017ApJ...834...85N}
{Newton}, E.~R., {Irwin}, J., {Charbonneau}, D., {et~al.} 2017, \apj, 834, 85

\bibitem[{{Oshagh}(2018)}]{2018ASSP...49..239O}
{Oshagh}, M. 2018, Asteroseismology and Exoplanets: Listening to the Stars and
  Searching for New Worlds, 49, 239

\bibitem[{{Pepe} {et~al.}(2010){Pepe}, {Cristiani}, {Rebolo Lopez}, {Santos},
  {Amorim}, {Avila}, {Benz}, {Bonifacio}, {Cabral}, {Carvas}, {Cirami},
  {Coelho}, {Comari}, {Coretti}, {De Caprio}, {Dekker}, {Delabre}, {Di
  Marcantonio}, {D'Odorico}, {Fleury}, {Garc{\'{\i}}a}, {Herreros Linares},
  {Hughes}, {Iwert}, {Lima}, {Lizon}, {Lo Curto}, {Lovis}, {Manescau},
  {Martins}, {M{\'e}gevand}, {Moitinho}, {Molaro}, {Monteiro}, {Monteiro},
  {Pasquini}, {Mordasini}, {Queloz}, {Rasilla}, {Rebord{\~a}o}, {Santana
  Tschudi}, {Santin}, {Sosnowska}, {Span{\`o}}, {Tenegi}, {Udry}, {Vanzella},
  {Viel}, {Zapatero Osorio}, \& {Zerbi}}]{EXPRESSO}
{Pepe}, F.~A., {Cristiani}, S., {Rebolo Lopez}, R., {et~al.} 2010, in
  \procspie, Vol. 7735, Ground-based and Airborne Instrumentation for Astronomy
  III, 77350F

\bibitem[{{Pizzolato} {et~al.}(2003){Pizzolato}, {Maggio}, {Micela},
  {Sciortino}, \& {Ventura}}]{2003A&A...397..147P}
{Pizzolato}, N., {Maggio}, A., {Micela}, G., {Sciortino}, S., \& {Ventura}, P.
  2003, \aap, 397, 147

\bibitem[{{Rajpaul} {et~al.}(2015){Rajpaul}, {Aigrain}, {Osborne}, {Reece}, \&
  {Roberts}}]{2015MNRAS.452.2269R}
{Rajpaul}, V., {Aigrain}, S., {Osborne}, M.~A., {Reece}, S., \& {Roberts}, S.
  2015, \mnras, 452, 2269

\bibitem[{{Rauer} {et~al.}(2016){Rauer}, {Aerts}, {Cabrera}, \& {PLATO
  Team}}]{2016AN....337..961R}
{Rauer}, H., {Aerts}, C., {Cabrera}, J., \& {PLATO Team}. 2016, Astronomische
  Nachrichten, 337, 961

\bibitem[{{Rauer} {et~al.}(2014){Rauer}, {Catala}, {Aerts}, {Appourchaux},
  {Benz}, {Brandeker}, {Christensen-Dalsgaard}, {Deleuil}, {Gizon}, {Goupil},
  {G{\"u}del}, {Janot-Pacheco}, {Mas-Hesse}, {Pagano}, {Piotto}, {Pollacco},
  {Santos}, {Smith}, {Su{\'a}rez}, {Szab{\'o}}, {Udry}, {Adibekyan}, {Alibert},
  {Almenara}, {Amaro-Seoane}, {Eiff}, {Asplund}, {Antonello}, {Barnes},
  {Baudin}, {Belkacem}, {Bergemann}, {Bihain}, {Birch}, {Bonfils}, {Boisse},
  {Bonomo}, {Borsa}, {Brand{\~a}o}, {Brocato}, {Brun}, {Burleigh}, {Burston},
  {Cabrera}, {Cassisi}, {Chaplin}, {Charpinet}, {Chiappini}, {Church},
  {Csizmadia}, {Cunha}, {Damasso}, {Davies}, {Deeg}, {D{\'{\i}}az}, {Dreizler},
  {Dreyer}, {Eggenberger}, {Ehrenreich}, {Eigm{\"u}ller}, {Erikson}, {Farmer},
  {Feltzing}, {de Oliveira Fialho}, {Figueira}, {Forveille}, {Fridlund},
  {Garc{\'{\i}}a}, {Giommi}, {Giuffrida}, {Godolt}, {Gomes da Silva},
  {Granzer}, {Grenfell}, {Grotsch-Noels}, {G{\"u}nther}, {Haswell}, {Hatzes},
  {H{\'e}brard}, {Hekker}, {Helled}, {Heng}, {Jenkins}, {Johansen},
  {Khodachenko}, {Kislyakova}, {Kley}, {Kolb}, {Krivova}, {Kupka}, {Lammer},
  {Lanza}, {Lebreton}, {Magrin}, {Marcos-Arenal}, {Marrese}, {Marques},
  {Martins}, {Mathis}, {Mathur}, {Messina}, {Miglio}, {Montalban}, {Montalto},
  {Monteiro}, {Moradi}, {Moravveji}, {Mordasini}, {Morel}, {Mortier},
  {Nascimbeni}, {Nelson}, {Nielsen}, {Noack}, {Norton}, {Ofir}, {Oshagh},
  {Ouazzani}, {P{\'a}pics}, {Parro}, {Petit}, {Plez}, {Poretti}, {Quirrenbach},
  {Ragazzoni}, {Raimondo}, {Rainer}, {Reese}, {Redmer}, {Reffert},
  {Rojas-Ayala}, {Roxburgh}, {Salmon}, {Santerne}, {Schneider}, {Schou},
  {Schuh}, {Schunker}, {Silva-Valio}, {Silvotti}, {Skillen}, {Snellen}, {Sohl},
  {Sousa}, {Sozzetti}, {Stello}, {Strassmeier}, {{\v S}vanda}, {Szab{\'o}},
  {Tkachenko}, {Valencia}, {Van Grootel}, {Vauclair}, {Ventura}, {Wagner},
  {Walton}, {Weingrill}, {Werner}, {Wheatley}, \&
  {Zwintz}}]{2014ExA....38..249R}
{Rauer}, H., {Catala}, C., {Aerts}, C., {et~al.} 2014, Experimental Astronomy,
  38, 249

\bibitem[{{Reiners}(2012)}]{2012LRSP....9....1R}
{Reiners}, A. 2012, Living Reviews in Solar Physics, 9, 1

\bibitem[{{Reiners} {et~al.}(2012){Reiners}, {Joshi}, \&
  {Goldman}}]{2012AJ....143...93R}
{Reiners}, A., {Joshi}, N., \& {Goldman}, B. 2012, \aj, 143, 93

\bibitem[{{Reiners} {et~al.}(2014){Reiners}, {Sch{\"u}ssler}, \&
  {Passegger}}]{2014ApJ...794..144R}
{Reiners}, A., {Sch{\"u}ssler}, M., \& {Passegger}, V.~M. 2014, \apj, 794, 144

\bibitem[{{Reinhold} {et~al.}(2019){Reinhold}, {Bell}, {Kuszlewicz}, {Hekker},
  \& {Shapiro}}]{2019A&A...621A..21R}
{Reinhold}, T., {Bell}, K.~J., {Kuszlewicz}, J., {Hekker}, S., \& {Shapiro},
  A.~I. 2019, \aap, 621, A21

\bibitem[{{Reinhold} {et~al.}(2013){Reinhold}, {Reiners}, \&
  {Basri}}]{2013A&A...560A...4R}
{Reinhold}, T., {Reiners}, A., \& {Basri}, G. 2013, \aap, 560, A4

\bibitem[{{Reinhold} {et~al.}(2020){Reinhold}, {Shapiro}, {Solanki}, {Krivova},
  {Cameron}, \& {Amazo-G\'{o}mez}}]{Reinhold_sub}
{Reinhold}, T., {Shapiro}, A.~I., {Solanki}, S.~K., {et~al.} 2020, Science,
  under revision

\bibitem[{{Ricker} {et~al.}(2015){Ricker}, {Winn}, {Vanderspek}, {Latham},
  {Bakos}, {Bean}, {Berta-Thompson}, {Brown}, {Buchhave}, {Butler}, {Butler},
  {Chaplin}, {Charbonneau}, {Christensen-Dalsgaard}, {Clampin}, {Deming},
  {Doty}, {De Lee}, {Dressing}, {Dunham}, {Endl}, {Fressin}, {Ge}, {Henning},
  {Holman}, {Howard}, {Ida}, {Jenkins}, {Jernigan}, {Johnson}, {Kaltenegger},
  {Kawai}, {Kjeldsen}, {Laughlin}, {Levine}, {Lin}, {Lissauer}, {MacQueen},
  {Marcy}, {McCullough}, {Morton}, {Narita}, {Paegert}, {Palle}, {Pepe},
  {Pepper}, {Quirrenbach}, {Rinehart}, {Sasselov}, {Sato}, {Seager},
  {Sozzetti}, {Stassun}, {Sullivan}, {Szentgyorgyi}, {Torres}, {Udry}, \&
  {Villasenor}}]{2015JATIS...1a4003R}
{Ricker}, G.~R., {Winn}, J.~N., {Vanderspek}, R., {et~al.} 2015, Journal of
  Astronomical Telescopes, Instruments, and Systems, 1, 014003

\bibitem[{{Roberts} {et~al.}(2012){Roberts}, {Osborne}, {Ebden}, {Reece},
  {Gibson}, \& {Aigrain}}]{2012RSPTA.37110550R}
{Roberts}, S., {Osborne}, M., {Ebden}, M., {et~al.} 2012, Philosophical
  Transactions of the Royal Society of London Series A, 371, 20110550

\bibitem[{{Roxburgh} {et~al.}(2007){Roxburgh}, {Catala}, \& {PLATO
  Consortium}}]{2007CoAst.150..357R}
{Roxburgh}, I., {Catala}, C., \& {PLATO Consortium}. 2007, Communications in
  Asteroseismology, 150, 357

\bibitem[{{Scargle}(1982)}]{1982ApJ...263..835S}
{Scargle}, J.~D. 1982, \apj, 263, 835

\bibitem[{{Shapiro} {et~al.}(2017){Shapiro}, {Solanki}, {Krivova}, {Cameron},
  {Yeo}, \& {Schmutz}}]{2017NatAs...1..612S}
{Shapiro}, A.~I., {Solanki}, S.~K., {Krivova}, N.~A., {et~al.} 2017, Nature
  Astronomy, 1, 612

\bibitem[{{Shapiro} {et~al.}(2016){Shapiro}, {Solanki}, {Krivova}, {Yeo}, \&
  {Schmutz}}]{Shapiroetal2016}
{Shapiro}, A.~I., {Solanki}, S.~K., {Krivova}, N.~A., {Yeo}, K.~L., \&
  {Schmutz}, W.~K. 2016, \aap, 589, A46

\bibitem[{{Shapiro, A. I.} {et~al.}(2020){Shapiro, A. I.}, {Amazo-G\'omez, E.
  M.}, {Krivova, N. A.}, \& {Solanki, S. K.}}]{paperI}
{Shapiro, A. I.}, {Amazo-G\'omez, E. M.}, {Krivova, N. A.}, \& {Solanki, S. K.}
  2020, A\&A, 633, A32.
\newblock \url{https://doi.org/10.1051/0004-6361/201936018}

\bibitem[{{Solanki} {et~al.}(2006){Solanki}, {Inhester}, \&
  {Sch{\"u}ssler}}]{2006RPPh...69..563S}
{Solanki}, S.~K., {Inhester}, B., \& {Sch{\"u}ssler}, M. 2006, Reports on
  Progress in Physics, 69, 563

\bibitem[{Solanki {et~al.}(2006)Solanki, Inhester, \&
  SchÃŒssler}]{Solanki_2006}
Solanki, S.~K., Inhester, B., \& SchÃŒssler, M. 2006, Reports on Progress in
  Physics, 69, 563.
\newblock \url{https://doi.org/10.1088%2F0034-4885%2F69%2F3%2Fr02}

\bibitem[{{Solanki} {et~al.}(2013){Solanki}, {Krivova}, \&
  {Haigh}}]{2013ARA&A..51..311S}
{Solanki}, S.~K., {Krivova}, N.~A., \& {Haigh}, J.~D. 2013, \araa, 51, 311

\bibitem[{{Spiegel} \& {Zahn}(1992)}]{1992A&A...265..106S}
{Spiegel}, E.~A., \& {Zahn}, J.~P. 1992, \aap, 265, 106

\bibitem[{{Torrence} \& {Compo}(1998)}]{1998BAMS...79...61T}
{Torrence}, C., \& {Compo}, G.~P. 1998, Bulletin of the American Meteorological
  Society, 79, 61

\bibitem[{{Ulrich}(1986)}]{1986ApJ...306L..37U}
{Ulrich}, R.~K. 1986, \apjl, 306, L37

\bibitem[{{Van Cleve} \& {Caldwell}(2016)}]{2016ksci.rept....1V}
{Van Cleve}, J.~E., \& {Caldwell}, D.~A. 2016, {Kepler Instrument Handbook},
  Tech. rep.

\bibitem[{{van Saders} {et~al.}(2018){van Saders}, {Pinsonneault}, \&
  {Barbieri}}]{2018arXiv180304971V}
{van Saders}, J.~L., {Pinsonneault}, M.~H., \& {Barbieri}, M. 2018, ArXiv
  e-prints, arXiv:1803.04971

\bibitem[{{Waldmeier}(1955)}]{Waldmeier1955}
{Waldmeier}, M. 1955, {Ergebnisse und Probleme der Sonnenforschung.} (KLeipzig,
  Geest \& Portig)

\bibitem[{{Walkowicz} \& {Basri}(2013)}]{Walkowicz2013}
{Walkowicz}, L.~M., \& {Basri}, G.~S. 2013, \mnras, 436, 1883

\bibitem[{{Wilson}(1965)}]{1965ApJ...142..773W}
{Wilson}, P.~R. 1965, \apj, 142, 773

\bibitem[{{Witzke} {et~al.}(2020){Witzke}, {Reinhold}, {Shapiro}, {Krivova}, \&
  {Solanki}}]{Witzke2020}
{Witzke}, V., {Reinhold}, T., {Shapiro}, A.~I., {Krivova}, N.~A., \& {Solanki},
  S.~K. 2020, arXiv e-prints, arXiv:2001.01934

\bibitem[{{Wright} \& {Drake}(2016)}]{2016Natur.535..526W}
{Wright}, N.~J., \& {Drake}, J.~J. 2016, \nat, 535, 526

\bibitem[{{Wright} {et~al.}(2011){Wright}, {Drake}, {Mamajek}, \&
  {Henry}}]{2011ApJ...743...48W}
{Wright}, N.~J., {Drake}, J.~J., {Mamajek}, E.~E., \& {Henry}, G.~W. 2011,
  \apj, 743, 48

\bibitem[{{Wright} {et~al.}(2018){Wright}, {Newton}, {Williams}, {Drake}, \&
  {Yadav}}]{2018MNRAS.479.2351W}
{Wright}, N.~J., {Newton}, E.~R., {Williams}, P. K.~G., {Drake}, J.~J., \&
  {Yadav}, R.~K. 2018, \mnras, 479, 2351

\bibitem[{{Yeo} {et~al.}(2014){Yeo}, {Krivova}, {Solanki}, \&
  {Glassmeier}}]{2014A&A...570A..85Y}
{Yeo}, K.~L., {Krivova}, N.~A., {Solanki}, S.~K., \& {Glassmeier}, K.~H. 2014,
  \aap, 570, A85

\bibitem[{{Zechmeister} \& {K{\"u}rster}(2009)}]{2009A&A...496..577Z}
{Zechmeister}, M., \& {K{\"u}rster}, M. 2009, \aap, 496, 577

\end{thebibliography}
 
\end{document}